\documentclass[toc]{PoS}

\usepackage{amssymb,cite}
\usepackage{amsmath}
\usepackage{amsfonts}
\usepackage{bbm}
\usepackage{amsthm}
\usepackage{mathrsfs}
\usepackage{color}

\usepackage{subcaption}

\usepackage{mathrsfs}
\usepackage{bm}
\usepackage[arrow,matrix,curve]{xy}

\usepackage{amstext,amscd,array}

\theoremstyle{plain}

\theoremstyle{definition}

\numberwithin{equation}{section}

\def\dd{\mathrm{d}}

\def\BBB{\mathscr{B}}

\newcommand{\obultimes}{\mathbin{\ooalign{$\otimes$\cr\hidewidth\raise0.17ex\hbox{$\scriptstyle\bullet\mkern4.48mu$}}}}
\newcommand{\ostartimes}{\mathbin{\ooalign{$\otimes$\cr\hidewidth\raise0.17ex\hbox{$\scriptstyle\star\mkern4.48mu$}}}}

\newcommand{\obulplus}{\mathbin{\ooalign{$\boxplus$\cr\hidewidth\raise0.295ex\hbox{$\scriptstyle\bullet\mkern4.7mu$}}}}

\usepackage{epsf}
\usepackage{epsfig}
\usepackage{wrapfig}

\input{epsf}

\newcommand{\mbf}[1]{{\boldsymbol {#1} }}
\def\ii{{\,{\rm i}\,}}
\def\dd{{\rm d}}

\newcommand{\CCH}{\mathscr{H}}

\newcommand{\calo}{\mathcal{O}}
\newcommand{\calp}{\mathcal{P}}

\newcommand{\eq}{\begin{equation}}
\newcommand{\eqend}{\end{equation}}
\newcommand{\eqa}{\begin{eqnarray}}
\newcommand{\nonueqa}{\begin{eqnarray*}}
\newcommand{\eqaend}{\end{eqnarray}}
\newcommand{\nonueqaend}{\end{eqnarray*}}

\newcommand{\bma}[1]{\begin{array}{#1}}
\newcommand{\ema}{\end{array}}
\newcommand{\bc}{\begin{center}}
\newcommand{\ec}{\end{center}}

\newcommand{\complex}{{\mathbb C}} 
\newcommand{\zed}{{\mathbb Z}} 
\newcommand{\real}{{\mathbb R}} 

\newcommand{\quat}{{\mathbb H}} 
\newcommand{\oct}{{\mathbb O}} 

\def\hil{{\mathcal H}}

\def\Mcal{{\mathcal M}}

\newif\ifold             \oldtrue

\def\e{{\,\rm e}\,}

\def\be{\begin{equation}}
\def\ee{\end{equation}}
\def\bea{\begin{eqnarray}}
\def\eea{\end{eqnarray}}
\def\bd{\begin{displaymath}}
\def\ed{\end{displaymath}}

\newcommand{\beq}{\begin{eqnarray}}
\newcommand{\eeq}{\end{eqnarray}}

\makeatletter
\newdimen\normalarrayskip              
\newdimen\minarrayskip                 
\normalarrayskip\baselineskip
\minarrayskip\jot
\newif\ifold             \oldtrue            
\def\arraymode{\ifold\relax\else\displaystyle\fi} 
\def\@arrayskip{\ifold\baselineskip\z@\lineskip\z@
     \else
     \baselineskip\minarrayskip\lineskip2\minarrayskip\fi}
\def\@arrayclassz{\ifcase \@lastchclass \@acolampacol \or
\@ampacol \or \or \or \@addamp \or
   \@acolampacol \or \@firstampfalse \@acol \fi
\edef\@preamble{\@preamble
  \ifcase \@chnum
     \hfil$\relax\arraymode\@sharp$\hfil
     \or $\relax\arraymode\@sharp$\hfil
     \or \hfil$\relax\arraymode\@sharp$\fi}}
\def\@array[#1]#2{\setbox\@arstrutbox=\hbox{\vrule
     height\arraystretch \ht\strutbox
     depth\arraystretch \dp\strutbox
     width\z@}\@mkpream{#2}\edef\@preamble{\halign \noexpand\@halignto
\bgroup \tabskip\z@ \@arstrut \@preamble \tabskip\z@ \cr}%
\let\@startpbox\@@startpbox \let\@endpbox\@@endpbox
  \if #1t\vtop \else \if#1b\vbox \else \vcenter \fi\fi
  \bgroup \let\par\relax
  \let\@sharp##\let\protect\relax
  \@arrayskip\@preamble}
\makeatother

\def\be{\beta}

\theoremstyle{definition}

\title{An Introduction to Nonassociative Physics}

\ShortTitle{An Introduction to Nonassociative Physics}

\author{\speaker{Richard J. Szabo}%
        \vspace{2mm}\\
Department of Mathematics, Heriot-Watt University, Edinburgh, United Kingdom.\vspace{1mm}\\
Maxwell Institute for Mathematical Sciences, Edinburgh, United Kingdom.\vspace{1mm}\\
The Higgs Centre for Theoretical Physics, Edinburgh, United Kingdom.\vspace{2mm}\\
E-mail: \email{R.J.Szabo@hw.ac.uk}}

\abstract{We give a pedagogical introduction to the nonassociative structures arising from recent developments in quantum mechanics with magnetic monopoles, in string theory and M-theory with non-geometric fluxes, and in M-theory with non-geometric Kaluza-Klein monopoles. After a brief overview of the main historical appearences of nonassociativity in quantum mechanics, string theory and M-theory, we provide a detailed account of the classical and quantum dynamics of electric charges in the backgrounds of various distributions of magnetic charge. We apply Born reciprocity to map this system to the phase space of closed strings propagating in $R$-flux backgrounds of string theory, and then describe the lift to the phase space of M2-branes in $R$-flux backgrounds of M-theory. Applying Born reciprocity maps this M-theory configuration to the phase space of M-waves probing a non-geometric Kaluza-Klein monopole background. These four perspective systems are unified by a covariant 3-algebra structure on the M-theory phase space.}

\FullConference{Corfu Summer Institute 2018 "School and Workshops on Elementary Particle Physics and Gravity"\\
		(CORFU2018)\\
		31 August--28 September 2018\\
		Corfu, Greece\\
{\small Preprint: \  {\tt EMPG--19--10}}}

\begin{document}

\newpage

\section{A brief history of nonassociativity in physics}

\subsection{Nonassociative algebra}

Nonassociative algebras first appeared around the middle of the
nineteenth century and have since formed an independent branch
of mathematics. By far the simplest and best known example of a
nonassociative algebra is a Lie algebra. A Lie bracket
$[\,\cdot\,,\,\cdot\,]$ on a vector
space is generally a noncommutative and
nonassociative binary operation in the sense that $[x,y]\neq[y,x]$ and
$[x,[y,z]]\neq
[[x,y],z]$, but rather it is antisymmetric and satisfies the
Jacobi identity. If we package the quantity which vanishes by the
Jacobi identity into a ternary operation called the `Jacobiator',
\bea\label{eq:Jacobiator}
{}[x,y,z] := \tfrac13\,\big([x,[y,z]]-[[x,y],z]+[[x,z],y]\big) \ ,
\eea
then Lie algebras are characterized by the feature that this 3-bracket
vanishes: $[x,y,z]=0$. In our present context this will be considered
to be a trivial example of a nonassociative algebra, as the binary
operation on the Lie group
which integrates the Lie algebra via the exponential map is associative. In fact, in this
paper we will be
looking at cases of algebras with a multiplication that has non-vanishing
Jacobiators, just like the multiplication in a noncommutative algebra can be characterized
by a non-vanishing commutator $[x,y]:=x\,y-y\,x\neq0$.
A well-known example of a noncommutative and nonassociative
algebra in this sense is the algebra of octonions, which constitutes
an important example of an
`alternative algebra', where associativity of
the multiplication (and hence the Jacobi identity) is generally violated, but it
possesses the alternativity condition which generalizes
the Jordan identity $(x^2\,y)\,x=x^2\,(y\,x)$ of (commutative) Jordan algebras,
which we review below. 

Algebras whose associativity is controlled by
identities like the Jacobi or alternative identities were of central mathematical interest in the
last century, as arbitrarily relaxing associativity in an algebra
does not lead to a good mathematical theory unless some other
structures are present, see e.g.~\cite{NA-Book}. They have recently made their way into the forefront
of some recent developments in quantum mechanics, string
theory and M-theory. The main aim of these lectures is to survey some of these
recent insights and to highlight some of the interesting physical
consequences of the theory, together with the many important open
avenues awaiting further investigation. 

Nonassociative algebras actually have a long and diverse history of appearences in
physics, which is perhaps not so widely appreciated. The purpose of this opening section is to briefly review some of
their occurences with a personally biased point of view: We only discuss developments in
physics with an eye to the main topics that we will pursue later
on, and to overview the developments that we shall treat in more
detail in subsequent sections.

\subsection{The octonions}
\label{sec:octonions}

For later use, let us begin by recalling the standard mathematical example of the octonions, as
it will play an important role in some of the discussions throughout
this paper (see e.g.~\cite{Baez:2001dm} for a nice introduction).
A well-known theorem in algebra states that there are only four normed
division algebras over the field $\real$ of real numbers: The real
numbers $\real$ themselves, the complex numbers $\complex$, the
quaternions $\quat$, and the octonions $\oct$. Both $\real$ and
$\complex$ are commutative and associative algebras, $\quat$ is
noncommutative but associative, while $\oct$ is both noncommutative
and nonassociative. The algebra $\real$ is generated over itself by a central
identity element $1$, $\complex$ is generated over $\real$ by $1$ and
an imaginary complex unit $\ii=\sqrt{-1}$, while $\quat$ is generated by $1$ and
three imaginary quaternion units $\ii$, ${\rm j}$ and ${\rm k}$, 
familar from quantum mechanics in their representation in terms of
Pauli spin matrices $e_i=\sqrt{-1}\, \sigma_i$, $i=1,2,3$, which obey $e_i^2=-1$
and
\bea
e_i\,e_j = -\delta_{ij} + \varepsilon_{ijk}\,e_k \ ,
\eea
where $\varepsilon_{ijk}$ is the Levi-Civita symbol in three
dimensions. 

Analogously, the algebra $\oct$ is
generated by $1$ and seven imaginary octonion units
$e_A$, $A=1,\dots,7$, satisfying $e_A^2=-1$, so that a generic element
$O$ of $\oct$ is of the form
\bea
O=a_0 +a_1\, e_1+a_2\,
    e_2\,+\cdots + a_7\, e_7 \ ,
\eea
with $a_0,a_1,\dots,a_7\in\real$. 
The multiplication in the algebra $\oct$ can be written in the form
\bea\label{eq:octonioncomm}
e_A\, e_B=-\delta_{AB}
  +\eta_{ABC}\, e_C  \ ,
\eea
where the structure constants $\eta_{ABC}$ form a completely
antisymmetric tensor with only seven non-vanishing components,
generalizing the structure constants $\varepsilon_{ijk}$ (with only
one non-zero component) of the algebra
of quaternions $\quat$. 

To specify $\eta_{ABC}$ explicitly, it is convenient to rewrite $e_4,e_5,e_6=f_1,f_2,f_3$
and represent the algebra of octonions in terms of the commutators
\begin{align}
{}[e_i,e_j]&={2\, \varepsilon_{ijk}\, e_k} \qquad \mbox{and} \qquad [e_7,e_i] = 2\,
             f_i \ , \nonumber \\[4pt]
{} [f_i,f_j]&=-2\, \varepsilon_{ijk}\, e_k \qquad \mbox{and} \qquad [e_7,f_i] =
             -2\, e_i \ , \\[4pt]
{} [e_i,f_j]&=2\, (\delta_{ij}\, e_7- \varepsilon_{ijk}\, f_k) \ . \nonumber 
\end{align}
The first set of commutation relations describes a quaternion
subalgebra $\quat\subset\oct$ (the Lie algebra of $SU(2)\simeq SO(3)$), analogous to the hierarchy of embeddings
$\real\subset\complex\subset\quat$ generated by the inclusions of $1$
and $\ii$; this follows from the Cayley-Dickson construction of the real
normed division algebras. This demonstrates the important feature that, unlike $\quat$ whose structure
constants are invariant under the full three-dimensional rotation
group $SO(3)$, the symmetry group of the tensor $\eta_{ABC}$ is only the
14-dimensional simple exceptional Lie subgroup $G_2\subset SO(7)$ of the rotation group in seven
dimensions. Using these commutators one
can derive the non-vanishing Jacobiators
\bea\label{eq:octonionjac}
[e_A,e_B,e_C]=-4\,\eta_{ABCD}\,e_D \ ,
\eea
with antisymmetric structure constants $\eta_{ABCD}$ also having only
seven non-vanishing components. The \emph{alternative} property of
the algebra $\oct$ is then equivalent to the statement that its Jacobiators
are proportional to its `associators':
\bea
[e_A,e_B,e_C]=
2\,\big((e_A\,e_B)\,e_C-e_A\,(e_B\,
e_C)\big) \ .
\eea

\subsection{Jordanian quantum mechanics\label{sec:Jordan}}

The algebra of octonions $\oct$ is an example of what may be called a
noncommutative Jordan algebra, which we now proceed to define. Jordan
algebras were the first appearence of nonassociativity
in physics, in the early days of quantum
theory~\cite{Jordan}. Jordan's motivation
for introducing them was to define an algebraic structure on
the set of observables in quantum
mechanics, and hence an overall new algebraic setting for the
foundations of quantum theory. The observables of any quantum
system are given by Hermitian elements in a $C^*$-algebra, which can
be represented by operators on a finite- or
infinite-dimensional separable Hilbert space. If $A$ and $B$ are Hermitian operators, then their operator product is not
a Hermitian operator (unless they commute), and neither is their
commutator. Thus neither of these binary operations closes on the set
of physical observables, and so they are not an intrinsic part of the 
physically meaningful characteristics of the system. Jordan
algebras were introduced to formalize the properties of a
finite-dimensional quantum system. 

The basic observation is that the symmetrized product
\bea
A\circ B = \tfrac12\,(A\,B+B\, A) = \tfrac12\,\big((A+B)^2-A^2-B^2\big)
\eea
\emph{is} Hermitian. This product is obviously commutative,
\bea
A\circ B =B\circ A \ ,
\eea
but it is nonassociative: $(A\circ B)\circ C\neq A\circ(B\circ C)$ in
general. However, it satisfies the \emph{Jordan identity}
\bea
(A^2\circ B)\circ A = A^2\circ(B\circ A)
\eea
which is equivalent to the \emph{power-associativity} property
\bea
A^n\circ A^m = A^{n+m} \ , 
\eea
for all non-negative integer powers $n$ and $m$. Hermitian
operators also have the property that $A^2+B^2=0$ implies $A=B=0$,
i.e. the symmetrized product on observables is \emph{formally real}.

Abstracting these properties leads to the notion of a \emph{Jordan
  algebra} as a commutative power-associative algebra; in other words,
although the set of observables do not form an algebra under the usual
multiplication of operators, they do form a Jordan algebra with
respect to the symmetrized product. If the multiplication on a Jordan algebra can be
brought via a suitable isomorphism to the form of a symmetrized product $\circ$ on an associative algebra,
as above, it is said to be \emph{special}. The
power-associative property implies that it is sufficient to demand
that the binary operation on a Jordan algebra is \emph{alternative}~\cite{Albert,Schafer},
\bea
(x\, y)\,x=x\, (y\, x) \ .
\eea
By abstracting this property we may thus define the notion of a
\emph{noncommutative Jordan algebra} wherein we relax the
commutativity of the multiplication.

The starting programme of Jordanian quantum mechanics was then to find examples of
non-special Jordan algebras as algebras of observables for quantum
systems. These hopes were dashed by the famous work of
Jordan, von~Neumann and Wigner~\cite{JvNW} who proved that the only
simple formally real finite-dimensional Jordan algebras are the Jordan algebras of $n \times
n$ Hermitian matrices over $\real$, $\complex$ and $\quat$, the
exceptional $27$-dimensional Albert algebra of $3 \times 3$ Hermitian
matrices over the octonions $\oct$, and the Clifford-type (or `spin') Jordan
algebras; among these, only the Albert algebra, equiped with the
symmetrized product, is non-special. An `octonionic quantum mechanics' based on the Albert
algebra was later developed by~\cite{Gunaydin:1978jq} which satisfies the
von~Neumann axioms of quantum theory, despite there being no Hilbert
space formulation for nonassociative algebras: Operators acting on a separable
Hilbert space necessarily associate. The Jordan-von~Neumann-Wigner theorem
was generalized quite some time later by
Zelmanov~\cite{Zelmanov} to the infinite-dimensional case. This
implies that there is no infinite-dimensional non-special Jordan
algebra that would accomodate the observables of quantum
mechanics, thus eliminating the prospects of Jordan algebras from
playing a viable role in the foundations of quantum theory.

\subsection{Nambu mechanics}
\label{sec:Nambu}

Let us now fast-forward ahead many years to a very different kind
of nonassociative structure that takes place directly at the level of
a ternary operation. Such structures were introduced by
Nambu~\cite{Nambu}; they were formalised and generalised 20 years later by
Takhtajan~\cite{Takhtajan:1993vr}. A \emph{Nambu-Poisson bracket} on a
manifold $M$ is a completely antisymmetric ternary operation
$\{\,\cdot\,,\,\cdot\,,\,\cdot\,\}$ on the algebra of functions on $M$
which
obeys the Leibniz rule,
\bea
\{f\,g,h,k\}=f\,\{g,h,k\}+\{f,h,k\}\,g \ ,
\eea
and a higher version of the Jacobi identity for Lie algebras called
the ``fundamental identity'',
\bea\label{eq:fundid}
\hspace{-5mm} \{f,g,\{h_1,h_2,h_3\}\} - \{\{f,g,h_1\},h_2,h_3\} +
\{\{f,g,h_2\},h_1,h_3\} - \{\{f,g,h_3\},h_1,h_2\} = 0 \ ,
\eea
for functions $f,g,h,h_1,h_2,h_3,k\in C^\infty(M)$. These axioms
extend the axioms of a Poisson bracket which defines a Lie algebra on
the vector space $C^\infty(M)$; likewise, a Nambu-Poisson bracket
defines a higher version of a Lie algebra on the vector space
$C^\infty(M)$ called a `3-Lie algebra'~\cite{Takhtajan:1993vr}. 

Nambu's idea was to reformulate the equations of motion of
complicated non-integrable dynamical systems as a bi-Hamiltonian dynamics with
flow equations 
\bea
\frac{\dd f}{\dd t} = \{f,H_1,H_2\}
\eea
involving
a pair of Hamiltonians $H_1$ and $H_2$. The hope was to discover new
(generalized) integrals of motion. Consistency of these time
evolution equations necessitates both the Leibniz rule and also the
fundamental identity if one requires time derivatives to act as
derivations of the 3-bracket, so that they play crucial roles
analogous to the roles of the Leibniz rule and Jacobi
identity of the Poisson bracket in ordinary Hamiltonian mechanics. 

Nambu's original example was
the canonical 3-bracket on $M=\real^3$ given by
\bea\label{eq:NPR3}
\{f,g,h\}:= \varepsilon^{ijk}\, \partial_i f\, \partial_j
g\, \partial_k h \ ,
\eea
and the motivating dynamical problem is given by
the Euler equations
\bea
\frac{\dd\vec L}{\dd t}=\vec L\,\mbf\times\,\vec\omega
\eea
describing the motion of a rotating rigid body in $\real^3$ with
orbital angular momentum $\vec L$ and angular velocity 
$\vec\omega$ about its principal axes. Nambu's observation was that
these equations are equivalent to
the bi-Hamiltonian equations
\bea
\frac{\dd L_i}{\dd t} = \big\{L_i,\vec L{\,}^2,T\big\}
\eea
for the components $L_i$ of the angular momentum with respect to the
3-bracket \eqref{eq:NPR3} with $\partial_i=\partial/\partial L_i$, and
the pair of Hamiltonians $H_1=\vec L\,^2$ 
and $H_2=T=\frac12\,\vec L\cdot \vec\omega$.

Attempts to quantize the Nambu-Poisson bracket brings us closer to our
earlier discussions of nonassociativity. If one wishes to use 
a correspondence principle analogous to the usual Bohr correspondence principle of quantum mechanics, then the
natural extension of quantizing Poisson brackets, by mapping them to commutators
of operators on a Hilbert space, would be to map
Nambu-Poisson brackets to Jacobiators, but this does not work
because operators on a separable Hilbert space always associate and so
their Jacobiators vanish, as mentioned before. Instead, one can
define the \emph{Nambu-Heisenberg bracket} to be
the ``half-Jacobiator''~\cite{Nambu}
\bea
[A,B,C]_\text{NH} = [A,B] \, C - [A,C] \, B + [B,C] \, A
\eea
consisting of an antisymmetric combination of half of all terms which
would otherwise vanish by the Jacobi
identity. This proposal unfortunately 
meets with several problems, most notably that the fundamental identity is not preserved under the correspondence
principle. There is a tension between the Leibniz rule and the
fundamental identity as, in contrast to the Jacobi identity satisfied
by a Poisson bracket, the
fundamental identity imposes algebraic in addition to differential constraints on
the allowed Nambu-Poisson brackets. Hence it is difficult to find a suitable 3-Lie algebra of
operators that would quantize the Nambu-Poisson 3-bracket of functions. In this sense the quantization of Nambu-Poisson brackets to
this day remains an open problem; see~\cite{DeBellis:2010pf} for a
more modern perspective on this problem within our present context, together with
further references. 

In light of these issues, Nambu suggested to use nonassociative
algebras to quantize his 3-bracket, which we can then do via the
Jacobiator defined in \eqref{eq:Jacobiator} instead of the less
natural half-Jacobiator.
These arguments thus demonstrate that the problem of quantizing
Nambu mechanics is intimately related to formulating
\emph{nonassociative quantum mechanics}. This perspective will be the
driving theme throughout the rest of this article.

\subsection{Nonassociativity in string theory}
\label{sec:NAstringtheory}

String theory has been known for a long time to lead to new notions of
geometry which have driven many developments and fields of
mathematics. As a candidate quantum theory of gravity, it also offers
a host of modifications and deformations of the notion of spacetime
itself. Nonassociative structures in string theory
have a relatively long history of appearences predating the more
modern developments that we shall discuss in subsequent sections, which
is again perhaps not so widely appreciated. 
Here we shall highlight
some of the most important selected examples of nonassociativity in the context of string theory.

\paragraph{Closed string field theory.}

Noncommutative geometry first made its appearence in string theory in
the mid-1980s with Witten's work on bosonic open string field
theory~\cite{Witten:1985cc}. Shortly thereafter, nonassociativity
originally appeared in string theory
through Strominger's work~\cite{Strominger:1987ad} which showed that the
algebraic structures of gauge symmetries
underlying bosonic closed string field theory do not possess the usual
algebraic features of conventional symmetries in quantum field theory
and string theory, which are governed infinitesimally by Lie algebras. The gauge symmetries of
the theory include local spacetime diffeomorphisms, and the
gauge-invariant action is constructed as the associator of Witten's
open string field star product. Whereas open string
states associate, the associator is non-vanishing on closed string states
due to anomalies in associativity~\cite{Horowitz:1987yz}, which nonetheless still leads to a
consistent theory. The complete analysis of the gauge symmetries
was carried out later on by Zwiebach~\cite{Zwiebach:1992ie} using the
Batalin-Vilkovisky formalism, who showed
that the full action of closed string field
theory is defined by an infinite chain of higher string field
products on the off-shell state space satisfying homotopy deformations of Jacobi-type
identities. Such a structure was identified as a strong homotopy Lie
algebra~\cite{Lada:1992wc,Stasheff:1993ny}, or $L_\infty$-algebra for
short, and has been realised
in the mathematics community to be the most natural and general
higher version of a Lie algebra which generically exhibits an infinite hierarchy
of nonassociative $n$-ary operations. See~\cite{Stasheff:2018vnl} for
a recent survey of the developments, together with both old and new applications in string theory.

\paragraph{D-branes in curved backgrounds.}

Witten's inception of noncommutative geometry in open string field
theory was reinvigorated in a much simpler and precise setting over ten years
later through the seminal work of Seiberg and
Witten~\cite{Seiberg:1999vs}. They showed that, in a particular
scaling limit which decouples the open and closed string modes, the worldvolume of a D-brane in the background of
a constant Kalb-Ramond field $B$ experiences a noncommutative
deformation of its geometry, and the low-energy effective field theory
on the D-brane is a
noncommutative version of Yang-Mills theory,  a non-local gauge
theory which retains stringy features in the framework of quantum field theory, such as an exact open string
T-duality invariance. This sparked the
flurry of exciting activity
in the physics community that has evolved into the area of research
dubbed `noncommutative field theory' (these days a subject class of many high
energy physics journals); see e.g.~\cite{Douglas:2001ba,Szabo:2001kg} for reviews of the accelerating
developments through the turn of the millenium. This development was
subsequently extended to the case of non-constant 2-form $B$-fields
in~\cite{Cornalba:2001sm,Herbst:2001ai}, which showed that the
noncommutative star products deforming the multiplication of fields on D-brane worldvolumes
become nonassociative, where the NS--NS 3-form flux $H=\dd B$ controls the
Jacobiators of star products. As in the case of string field theory,
nonassociativity in this case is an \emph{off-shell} phenomenon: In
the integrated on-shell open string scattering amplitudes the effects
of nonassociativity (but not noncommutativity) vanish due to
appropriate cyclicity properties of the nonassociative
star products~\cite{Herbst:2003we}, consistently with the requirements
of crossing symmetry of the worldsheet conformal field theory of open
strings (which still involves ordinary quantum fields that necessarily
associate); see e.g.~\cite{Szabo:2006wx} for a review of these and other developments
along these lines. This physical caveat will reappear in various contexts of our
discussions of nonassociativity in subsequent sections. The gauge symmetries
underlying these and other nonassociative Yang-Mills theories arising
as decoupling limits of open
string theory also do not close in the conventional way to Lie algebras, but have been recently
shown to naturally form $L_\infty$-algebras
in~\cite{Blumenhagen:2018kwq}.

\paragraph{Topological T-duality.}

Around the middle of the last decade a mathematically rigorous
approach to T-duality was developed for principal torus bundles, and
also more general backgrounds with abelian
isometries~\cite{Mathai:2004qq,Bouwknegt:2004ap,Brodzki:2006fi}. The
algebra of functions in an NS--NS
$H$-flux background can be described by a mildly noncommutative `continuous trace algebra' which
incorporates the effects of $H$-flux as the Dixmier-Douady
characteristic class of a gerbe, just like a Kalb-Ramond field $B$
with $\dd B=0$ can be regarded as a background magnetic field, and `topological T-duality' can be
defined in a purely algebraic way on these functions. It
was realised that when the legs of the $H$-flux are aligned in
a particular way along the fibre directions on which T-duality
transformations are performed, the algebra of functions can map to noncommutative and even
nonassociative algebras in the T-dual background. However, the phrase `topological T-duality' refers to the
fact that such algebraic T-duality transformations are not the same as
the geometric maps found in the string theory literature. In
particular, one cannot recover the conventional commutative algebra of
functions on the fibration when $H$-flux is turned off; rather one
only recovers functions up to `Morita equivalence', which is a
``symmetry'' (in the physical sense) only insofar as it provides a
natural isomorphism between the corresponding K-theory groups. This was the original
motivation in viewing topological T-duality as a weakened version of open string
T-duality, as these K-theory groups classify the charges of D-branes
in the given backgrounds. It was, however, suggested that the
phenomenon should be more general and also apply to the closed string
sector, i.e. not only to D-brane
worldvolumes, but also to the background spacetime itself, and such algebras were
argued to be global versions of the `non-geometric' string backgrounds
which have no description in conventional Riemannian
geometry. These deformations of
closed string theory have
reemerged in more recent years in a somewhat different way which avoids the
unphysical features of Morita equivalence. They will constitute some
of our primary examples in subsequent sections and we shall review them in
more detail there.

\subsection{Nonassociativity in M-theory}
\label{sec:NAMtheory}

M-theory is the as yet unknown 11-dimensional quantum theory which unifies
all of the consistent 10-dimensional superstring theories, and it has
likewise suggested many classical and quantum deformations of geometry.
Nonassociative structures have also appeared independently and in a
sense more naturally in M-theory
in somewhat different fashions, also predating the more modern
developments we discuss in the following, whose most important highlights we
summarise here.

\paragraph{Multiple M2-branes.}

In the last decade substantial progress was made in understanding the
quantum geometry probed by the membranes of M-theory, which lift the
fundamental string degrees of freedom from type~IIA string
theory. This began with the work of Basu and Harvey~\cite{Basu:2004ed}
who proposed a lift of the D1--D3-brane system in string theory to the
M2--M5-brane system in M-theory. A BPS magnetic monopole solution of maximally
supersymmetric Yang-Mills theory in $3{+}1$ dimensions can be embedded into
string theory as a collection of open D1-branes
ending on a D3-brane which polarize into fuzzy 2-spheres, obtained as the
well-known
quantization of the canonical Poisson bracket on $S^2$. In the lift to
M-theory, this describes a self-dual string
wherein open M2-branes ending on an M5-brane polarize into
fuzzy 3-spheres obtained as a putative quantization of the canonical
Nambu-Poisson 3-bracket on $S^3$; this latter quantum geometry is
still not fully understood, see~\cite{DeBellis:2010pf} for a related
discussion and some partial attempts. As an attempt to lift the
nonabelian gauge theory which governs the low-energy description of
multiple D2-branes in string theory, the famous Bagger-Lambert theory
then generalized this 3-bracket structure which was
proposed to underlie a nonassocative algebra in which
the fields of a theory of multiple M2-branes take
values~\cite{Bagger:2006sk} and it was subsequently gauged to govern higher
gauge symmetries in these theories~\cite{Bagger:2007jr}, initially in the form of a 3-Lie algebra
(see e.g.~\cite{Bagger2012} for a review with more background and further
references). This appearence of 3-algebras constitutes the
first explicit appearence of nonassociativity in M-theory in the sense
discussed in Section~\ref{sec:Nambu}; the relation between the
3-algebras occuring in the Bagger-Lambert model and the quantization of
Nambu-Poisson brackets was explored in~\cite{DeBellis:2010sy} through
the framework of reduced models, for which explicit exact localization
computations were developed
in~\cite{DeBellis:2013zva}. The relation between these 3-algebras and
$L_\infty$-algebras is reviewed in e.g.~\cite{Saemann:2016sis}.

\paragraph{M5-branes in $\boldsymbol C$-field backgrounds.}

The natural lift to M-theory of the situation in which open strings
ending on a D2-brane, in the background of a constant NS--NS 2-form
$B$-field of 10-dimensional supergravity,
is the scenario in which open M2-branes end on an M5-brane, in the
background of a constant 3-form $C$-field of 11-dimensional
supergravity. Whereas the boundaries of open strings are point
particles and hence probe a noncommutative worldvolume theory on a
D2-brane, the boundaries of M2-branes are closed strings and hence are
expected to probe a noncommutative geometry on the \emph{loop space}
of the M5-brane worldvolume. These expectations were confirmed
explicitly in~\cite{Bergshoeff:2000jn,Kawamoto:2000zt}, which worked out a
somewhat complicated noncommutative deformation of loop space. It was
proposed almost 10 years later, using the Basu-Harvey equation, that the noncommutative
geometry probed by open membranes on the loop space can be more
compactly and simply understood in terms of a direct
nonassociative 3-algebra deformation of the M5-brane worldvolume
geometry itself, akin to what appears in the
Bagger-Lambert theory, and that this 3-algebra should
arise from a quantization of the Nambu-Poisson 3-bracket associated to
the constant $C$-field~\cite{Chu:2009iv}, in the same way that the
noncommutative geometry on a D2-brane arises as a quantization of the
Poisson bracket associated to a constant $B$-field. The connection between these
noncommutative loop space and nonassociative worldvolume 3-algebra structures was
established precisely in~\cite{Saemann2011,Saemann13} using transgression techniques; we
will comment further about this associative loop space perspective on
nonassociativity later on. Nambu-Poisson brackets, and their
quantization, have appeared in various other contexts in M-theory
since its inception in the mid-1990s,
see e.g.~\cite{Ho:2016hob} for a review.

\paragraph{${\mbf{G_2}}$ and ${\mbf{Spin(7)}}$ backgrounds.}

A somewhat more indirect and subtle appearence of nonassociativity in M-theory
has emerged over the years through the well-known compactifications of 11-dimensional
supergravity on Riemannian manifolds with holonomy valued in the groups $G_2$ or
$Spin(7)$, which are real Ricci-flat manifolds, see
e.g.~\cite{Acharya:2002gu,Gukov:2001hf}. Recall from Section~\ref{sec:octonions} that the
exceptional Lie group $G_2\subset SO(7)$ is precisely the automorphism group of
the algebra of octonions $\mathbb{O}$. A $G_2$-structure on a seven-dimensional
manifold is defined locally as a 3-form using the structure constants
$\eta_{ABC}$ of the commutator \eqref{eq:octonioncomm} of octonions which specify a
nonassociative binary operation. Likewise, the structure constants
$\eta_{ABCD}$ of the Jacobiator
\eqref{eq:octonionjac} combine with $\eta_{ABC}$ to locally define a self-dual
4-form which gives a reduction of the structure group of the oriented frame
bundle of an eight-dimensional manifold to $Spin(7)\subset SO(8)$. For
a detailed introduction, see e.g.~\cite{Salamon:2010cs}. These $G$-structures
will reappear in some of our later discussions of
non-geometry in M-theory, though in
a somewhat different context which relates them directly to
non-geometric string theory.
 
\subsection{Outline}

Let us now come to an overview of the main topics that we shall discuss in more
detail in the following. In these lectures, we will focus on two intimately
related occurences of nonassociativity in physics which have recently
received intensive investigation:

\paragraph{Magnetic monopoles.}

A pedagogical but rich example of nonassociative quantum mechanics,
unrelated to the Jordanian framework, arises in the quantization of
the dynamics of electric charges moving in the background of fields of magnetic charge. This observation is in itself not
recent and dates back to the
mid-1980s~\cite{Jackiw1985,Gunaydin:1985ur}. This work was
inspired at the time in part to understand, in a simpler framework,
the commutator anomalies that arise in certain
gauge theories coupled to chiral fermions which lead to violations of
the Jacobi identity among the currents generating local gauge
symmetries, see
e.g.~\cite{Faddeev:1985iz,Faddeev:1986pc,Semenoff:1987xw}. These
models also provide much simpler examples, at the level of quantum
mechanics and quantum field theory, of the associativity
anomalies that we discussed in Section~\ref{sec:NAstringtheory} which
were encountered in closed string field theory at around the same
time. As in the setting of string field theory, it was argued that
these instances still lead to sensible quantum theories when suitably
interpreted. Recent developments have gone beyond these
interpretations to illustrate that a nonassociative version of quantum
mechanics is not only physically sensible, but also has the potential to be
experimentally detected. This is the content of
Section~\ref{sec:magcharge}.

\paragraph{Locally non-geometric fluxes.}

In the course of this decade it has been suggested that closed
strings should probe a noncommutative and
nonassociative deformation of spacetime geometry in order to capture a
global perspective on
locally non-geometric flux compactifications of string theory, as originally
suggested by~\cite{Brodzki:2007hg}, at least for those which are obtained via T-duality
from geometric backgrounds with an NS--NS $H$-flux; this extends, to the closed
string sector, the
well-established case of open strings probing a noncommutative (but (on-shell)
associative) deformation of the worldvolume geometry on a D-brane in
the presence of a Kalb-Ramond field. This suggestion has by now been
interpreted from many distinct points of view, see
e.g.~\cite{stnag1,stnag2,Blumenhagen2011,Mylonas2012,Chatzistavrakidis2013,Blumenhagen:2013zpa,Blair:2014kla,Freidel:2017nhg}. This
is the content of Section~\ref{sec:stringsflux}.

The locally non-geometric backgrounds of string theory have lifts to
M-theory, where it is purported that closed M2-branes probing a
noncommutative and nonassociative deformation of spacetime geometry
also capture locally non-geometric backgrounds, at least for
four-dimensional compactifications of
M-theory~\cite{Gunaydin:2016axc,Kupriyanov:2017oob}. These
nonassociative 
structures are intimately related to our discussion of the octonions
from Section~\ref{sec:octonions}, and also to our discussion of the
role of $G_2$- and $Spin(7)$-holonomy manifolds in compactifications
of M-theory from
Section~\ref{sec:NAMtheory}; they are the topic of
Section~\ref{sec:M2flux}.

The noncommutative geometry of D3-branes in a constant $B$-field
background is completely analogous to that of the motion of electric
charges in constant magnetic fields; see e.g.~\cite{Szabo:2004ic} for an
introduction. Likewise, the nonassociative geometry of non-geometric string backgrounds can
also be understood from a certain ``dual'' perspective, in a
sense explained below, which relates them to the electron-monopole
system discussed above. An electric charge propagating in a distribution of
magnetic monopoles can be
embedded into string theory as a D0-brane probing a distribution of
D6-branes, giving another perspective on the nonassociative
deformation of the spacetime transverse to the worldvolume of the D6-branes. The lift of this D0--D6-brane system
to M-theory is an M-wave probing a non-geometric Kaluza-Klein monopole
background of M-theory~\cite{Lust:2017bgx}, which ties all of our three main
examples from Sections~\ref{sec:magcharge}--\ref{sec:M2flux} together,
with profound physical consequences, and is the topic of
Section~\ref{sec:MWflux}.

\paragraph{ \ }

These lecture notes are pedagogically written and unapologetically geared
at a general physics audience,
without assuming any detailed requisite technical knowledge of non-geometric backgrounds
or of the interesting but sophisticated mathematics of the higher structures
involved. A recent review of non-geometric backgrounds most relevant
to our current exposition can be found
in~\cite{Plauschinn:2018wbo}. More expository technical details on the
quantization of non-geometric backgrounds, their relation to double
field theory and the higher structures involved, as well as of nonassociative
quantum mechanics in these contexts are found in the lecture
notes~\cite{Szabo:2018hhh}. The more mathematically inclined reader
interested in higher structures may consult the mathematical
introduction~\cite{SzaboDurham} to
the ideas presented in the following. We will give further pointers to relevant
literature as we move along.

\section{Electric charges in magnetic monopole backgrounds}
\label{sec:magcharge}

\subsection{Magnetic Poisson brackets}
\label{sec:magbrackets}

The purpose of this section is to describe a very simple example of
nonassociativity which arises in the classical and quantum dynamics
of an electric charge in fields of magnetic charge. But we begin by
descibing the underlying geometry of the kinematical problem in a general way which
is also tailored to our later discussions of non-geometric string theory.

We work on the $d$-dimensional configuration space $M=\real^d$ whose
phase space $\Mcal=T^*M$ is equiped with local Darboux coordinates $X^I=(x^i,p_i)$,
where $x^i$ are coordinates on $M$ and $p_i$ are (cotangent) momentum
coordinates. The canonical Poisson brackets on $\Mcal$ are defined in
the usual way by the inverse of the canonical symplectic 2-form
\bea\label{eq:Darboux2form}
\omega_0=\dd p_i\wedge\dd x^i \ .
\eea
Any given 2-form $B\in\Omega^2(M)$,
not necessarily closed, will be refered to generally as a `magnetic
field' in what follows, for reasons that will become clear soon. The
magnetic field twists the symplectic form $\omega_0$ to an almost
symplectic form
\bea\label{eq:omegaB}
\omega_B=\omega_0-B \ ,
\eea
where the adjective `almost' refers to the fact that $\omega_B$ is
nondegenerate (because $\omega_0$ is) but not
necessarily closed: $\dd\omega_B=0$ if and only if $\dd B=0$. Its
inverse $\theta_B=\omega_B^{-1}$ defines the {magnetic Poisson
brackets} 
\bea\label{eq:magneticPoisson}
\{f,g\}_B = \theta_B^{IJ}\, \partial_If\,\partial_Jg
\eea
on functions $f,g\in C^\infty(\Mcal)$. On the local coordinates of
phase space these brackets read as 
\bea\label{eq:xpmagPoisson}
\{x^i,x^j\}_B=0 \ , \quad \{x^i,p_j\}_B=\delta^i{}_j \qquad \mbox{and}
\qquad \{p_i,p_j\}_B = -B_{ij}(x) \ ,
\eea
and they provide a deformation of the canonical Poisson brackets on phase
space. The dynamical model based on these magnetic Poisson brackets is
a generalization of the G\"unaydin-Zumino model~\cite{Gunaydin:1985ur}.

The noteworthy feature here is that these brackets only fulfill the
Jacobi identity $\{f,g,h\}_B=0$ for all $f,g,h\in C^\infty(\Mcal)$ when $\omega_B$ is a symplectic 2-form, i.e. when the
magnetic field is closed, $\dd B=0$. In general, we may call the
3-form $H=\dd B$ a `magnetic charge', also for reasons that will
become apparent very soon, and it controls the associativity of the
magnetic Poisson brackets \eqref{eq:magneticPoisson} on the phase
space $\Mcal$; in particular, on the local coordinates one finds that
only triples of momenta can nonassociate with the nonvanishing
Jacobiators
\bea
\{p_i,p_j,p_k\}_B=-H_{ijk}(x) \ .
\eea
In Poisson geometry one would thereby refer to the brackets
\eqref{eq:magneticPoisson} as \emph{$H$-twisted Poisson
  brackets}~\cite{Klimcik:2001vg,Severa:2001qm}.

\subsection{Magnetic monopoles}
\label{sec:monopoles}

One of the main motivations for the general definition of magnetic Poisson brackets
from Section~\ref{sec:magbrackets} is that they govern the Hamiltonian
dynamics of
electrically charged particles in backgrounds of magnetic
charge in classical and quantum mechanics. This is the case of $d=3$
dimensions wherein the 2-form $B$ can be written
using Poincar\'e duality in terms of a vector field $\vec B$ in
components as $B_{ij}=e\,\varepsilon_{ijk}\, B^k$. In this case the
magnetic Poisson brackets govern the motion of an electric charge $e$
in a static magnetic field $\vec B$ on $\real^3$. There are three
distinct cases to consider, all of which will be described in some
detail in the following.

\paragraph{Maxwell theory.}

The associative case where $\dd B=0$ corresponds in three dimensions
to the classical Maxwell theory of electromagnetism, in which there
are no magnetic monopoles. In terms of the magnetic field $\vec B$ this
case reads as
\bea\label{eq:divfree}
\vec\nabla\cdot\vec B=0 \ ,
\eea
and on $M=\real^3$ it implies the existence of a globally defined
magnetic vector potential $\vec A$ with
\bea
\vec B=\vec\nabla\,\mbf\times\, \vec A \ ,
\eea
or in general dimensions $d$ that $B=\dd A$ for a globally defined
1-form $A$.

\paragraph{Dirac monopoles.}

Dirac's semiclassical modification of Maxwell's theory postulates a
magnetic field $\vec B_g$ on $\real^3$ which is sourced by a
point magnetic charge $g$ at the origin of $\real^3$. This modifies
the Maxwell equation \eqref{eq:divfree} to
\bea
\vec\nabla\cdot\vec B_g=4\pi\,g \, \delta^{(3)}(\vec x\,) \ .
\eea
This describes a singular distribution of magnetic charge which, by
removing the location of the magnetic monopole, can be solved by the
magnetic field on $M^\times= \real^3\setminus\{\,\vec0\,\}$ given by
\bea\label{eq:Diracfield}
\vec B_g=g\,\frac{\vec x}{|\vec x\,|^3} \ .
\eea
On $M^\times$ one can write the Dirac monopole field in terms of a
locally defined magnetic vector potential $\vec A_{g,\vec n}$,
\bea\label{eq:Diracpotential}
\vec B_g=\vec\nabla\,\mbf\times\,\vec A_{g,\vec n} \qquad \mbox{with} \quad
\vec A_{g,\vec n}=\frac{g}{|\vec x\,|} \, \frac{\vec x\,\mbf\times\,\vec n}{|\vec
  x\,|-\vec x\cdot \vec n} \ ,
\eea
which has additional singularities along the infinite line in the
direction of a fixed unit vector $\vec n$, which is the celebrated
Dirac string singularity. Note that the Jacobiator of the magnetic
Poisson brackets in this case vanishes on $M^\times$.

\begin{figure}[htb]
\begin{center}
\includegraphics[width=5cm]{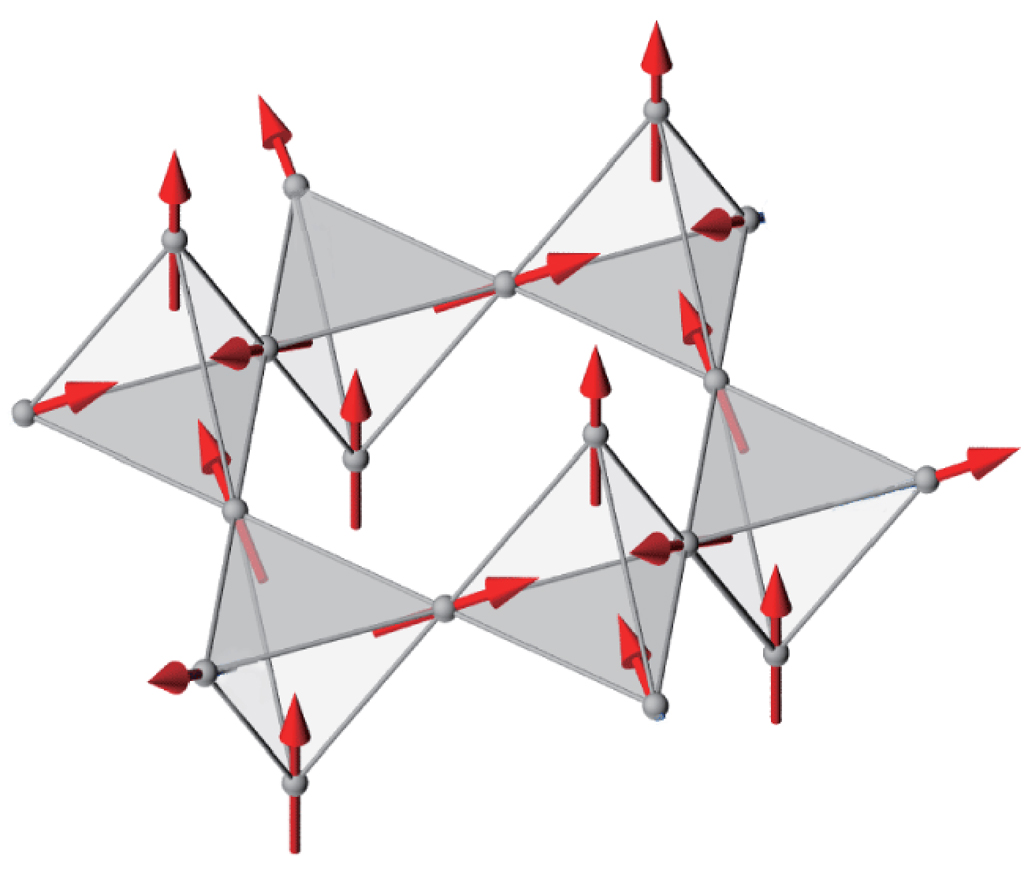}
\end{center}
\caption{\label{fig:spinice}\small Spin ice pyrochlore lattice with magnetic dipoles.}
\end{figure}

Magnetic monopoles are hypothetical particles with a single magnetic
pole whose existence would explain the quantization of electric
charge, as we discuss below. In spite of their vast interest in
theoretical physics, they have not yet been 
observed with full success in experiment. They were first observed in
analog condensed matter systems over ten
years ago where they appear as emergent states of
matter~\cite{Castelnovo:2007qi,Morris:2010ma}. In these experiments,
certain rare earth elements with the structure of a spin ice
pyrocholore lattice, consisting of an array of tetrahedra with
magnetic dipoles arranged at the corner atoms such that the total
magnetic charge through each tetrahedron vanishes (Figure~\ref{fig:spinice}), are
subjected to an external magnetic field and scatter neutrons, which
have their own magnetic dipole moment. Local configurations of
monopoles form and the resulting Dirac strings can be observed through
the neutron scattering interference patterns; see~\cite{Szabo:2017yxd}
for further details in the context of this paper together with more
references to the most important experiments. 

\paragraph{Smooth sources of magnetic charge.}

In this article we are primarily interested in the case of
non-singular monopole distributions, wherein
$\vec\nabla\cdot\vec B\neq0$ is a smooth function, or in general
dimensions $d$ the 3-form $H=\dd B\neq0$ is smooth; in this instance
the Jacobiator of the magnetic Poisson brackets in non-vanishing
throughout the smooth submanifold supporting the magnetic charge distribution. These are the
cases of interest in our later discussions of nonassociativity in
string theory and M-theory, but we shall also describe other potential
physical and even observable consequences of such models. We shall be
particularly interested in the cases of uniform distributions of
magnetic charge, for which $\vec\nabla\cdot\vec B=\rho$ is constant,
as these are the cases which are typically amenable to exact analytic
solution; then nonassociativity persists throughout all space. 

\subsection{Classical motion in fields of magnetic charge\label{sec:classical}}
 
Although we are primarily interested in the quantum mechanics of
electric charge in monopole backgrounds, we can gain a lot of
physical intuition about the qualitative differences between the different types of sources
of magnetic charge, discussed in Section~\ref{sec:monopoles}, by first
discussing the classical mechanics. In $d=3$ dimensions, the motion of
an electric charge $e$ with mass $m$ in the background of a static magnetic field $\vec B$ (with or
without sources) is governed by the Lorentz force law
\bea\label{eq:Lorentzforce}
\frac{\dd\vec p}{\dd t}=\frac em\,\vec p\,\mbf\times\, \vec B 
\eea
for the kinematical momentum 
\bea\label{eq:kinmom}
\vec p=m\,\frac{\dd\vec x}{\dd t}
\eea
of the electron. We ignore the
effects of magnetic backreaction due to the acceleration of the
electric charge. Using the
magnetic Poisson brackets \eqref{eq:xpmagPoisson}, these equations of
motion are equivalent to Hamilton's equations 
\bea
\frac{\dd X^I}{\dd t}=\big\{X^I,{\sf H}\big\}_B
\eea
for the Hamiltonian
\bea\label{eq:classHam}
{\sf H}=\frac1{2m}\, \vec p\,^2 \ .
\eea
The particular behaviour of the trajectories determined by these
dynamical evolution equations depends drastically on the nature of the
magnetic field $\vec B$.

\paragraph{Uniform magnetic fields.}

\begin{figure}[htb]
\begin{center}
\begin{subfigure}{.3\textwidth}
  \centering
  \includegraphics[width=.5\linewidth]{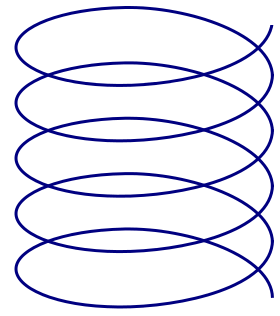}
  \caption{\small }
  \label{fig:sfig1}
\end{subfigure}%
\begin{subfigure}{.3\textwidth}
  \centering
  \includegraphics[width=.6\linewidth]{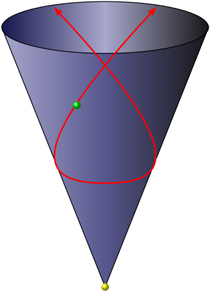}
  \caption{\small }
  \label{fig:sfig2}
\end{subfigure}
\begin{subfigure}{.3\textwidth}
  \centering
  \includegraphics[width=.3\linewidth]{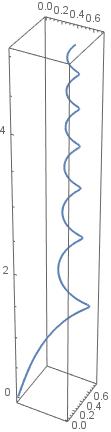}
  \caption{\small }
  \label{fig:sfig3}
\end{subfigure}
\caption{\small Classical electron trajectories in the background of: (a) a
  uniform magnetic field; (b) a Dirac monopole field; and (c) an axial
magnetic field.}
\label{fig:sfig}
\end{center}
\end{figure}

The textbook example of a source-free background with $\vec\nabla\cdot\vec B=0$
is given by a constant magnetic field $\vec B$ on $\real^3$. In this case the classical dynamics is
integrable and the motion is uniform along the direction of the
magnetic field. The equations of motion reduce to those of a
two-dimensional harmonic
oscillator, with angular frequency the Larmor frequency $e\,|\vec B\,|/m$, in the plane
perpendicular to $\vec B$ where the orbits are circular trajectories. Thus the motion follows a helical
trajectory with uniform velocity along the direction of the magnetic
field, as depicted in Figure~\ref{fig:sfig1}.

\paragraph{Dirac monopole fields.}

In the background of the field $\vec B=\vec B_g$ of a single Dirac
monopole \eqref{eq:Diracfield}, the classical dynamics is also
integrable. The integrals of motion are provided by the Poincar\'e
vector $\vec K$, which is the sum of the orbital angular momentum
$\vec L$ with the angular momentum
of the electromagnetic field due to the electric charge and the Dirac
monopole. The conservation of the Poincar\'e vector confines the
motion of the electron to the surface of a cone, with apex at the
location of the monopole, which precesses around the direction $\vec
K$ with a time-dependent angular frequency as depicted in
Figure~\ref{fig:sfig2}. In this case the 
electric charge never reaches the magnetic monopole, so the motion
takes place in $M^\times$ and
nonassociativity of the magnetic Poisson brackets plays no
role~\cite{BL}. This dynamical system was generalized recently to the case of
multiple electric charges and monopoles 
in~\cite{Heninger:2018guf}, where it was shown that the Jacobiator of
the corresponding magnetic Poisson brackets vanishes only if all
charges have identical ratio of electric to magnetic charge, or if an
electron never collides with a monopole.

\paragraph{Axial magnetic fields.}

In the case of a uniform distribution of magnetic charge $\rho$, the
closest analog to the case of a uniform magnetic field considered
above is when the magnetic field is aligned along the
direction of a fixed unit vector $\vec n$ in $\real^3$, so that $\vec
B=\vec B_{\rho,\vec n}$ with
\bea
\vec B_{\rho,\vec n} = \rho \, (\vec x\cdot\vec n) \ \vec n \ .
\eea
The corresponding Lorentz force equation was studied in~\cite{Kupriyanov:2018xji}, where
it was shown that the motion is again completely integrable. Again the
motion is uniform along the direction $\vec n$ of the magnetic field,
and in the plane perpendicular to $\vec n$ the equations of motion are
those of a harmonic oscillator with a time-dependent angular frequency, which
can be solved in terms of Fresnal integrals. In this case the
trajectories follow an Euler spiral with uniform velocity along the
direction $\vec n$ of the magnetic field, as depicted in
Figure~\ref{fig:sfig3}.

\paragraph{Rotationally symmetric magnetic fields.}

The situation is drastically different in the case of a uniform
distribution of magnetic charge $\rho$ which sources the
rotationally symmetric magnetic field $\vec B=\vec B_{\rho,{\rm rot}}$
with
\bea\label{eq:Brot}
\vec B_{\rho,{\rm rot}} = \tfrac13\, \rho \, \vec x \ .
\eea
In~\cite{BL} it was shown that the Lorentz force equation can be
transformed via a time reparameterization of the trajectories into the equation of motion
of a dissipative system, which governs the motion of an electric
charge in the field of a Dirac monopole $\vec B_g$ with an additional time-dependent
frictional force accounting for the background magnetic charge
$\rho$. In this case, the motion is no longer confined in any
direction, and the dynamics is not integrable. We shall discuss
another interpretation of this dissipative nature of the dynamics
later on.

\paragraph{ \ }

We shall now turn to the quantization of the dynamics and ask, in the
case of distributions of magnetic charge, if there is a sensible
version of nonassociative quantum mechanics accompanying the classical
analysis above.

\subsection{Quantization of magnetic Poisson brackets\label{sec:quantization}}

Let us examine the general problem of quantizing the magnetic Poisson brackets
\eqref{eq:magneticPoisson} on $M=\real^d$. Quantization
can be generally regarded as
a one-to-one linear map $f\mapsto\calo_f$ from functions $f\in C^\infty(\Mcal)$ to
symbols $\calo_f$ which admit some binary composition
operation $\circ$ with unit; the precise specification of the operators $\calo_f$ is
intentionally left vague for the moment, as this will depend crucially
on the quantization scheme chosen. The map $\calo$ should satisfy
the algebraic property
\bea
\lim_{\hbar\to0} \ \frac12 \,
\calo^{-1}\big(\calo_f\circ\calo_g+\calo_g\circ\calo_f\big) =
f\, g \ ,
\eea
where $f\,g$ is the usual pointwise multiplication of phase space
functions $f,g\in C^\infty(\Mcal)$, together with the generalized Bohr
correspondence principle
\bea
\lim_{\hbar\to0} \ \frac\ii\hbar \, 
\calo^{-1}\big([\calo_f,\calo_g]_\circ\big) = \{f,g\}_B \ ,
\eea
where
$[\calo_f,\calo_g]_\circ=\calo_f\circ\calo_g - \calo_g\circ\calo_f$. 
This is a generalization of Dirac's quantization framework which
deforms the algebraic and geometric structures of the classical phase
space $\Mcal$ in terms of some parameter $\hbar$, physically understood
as Planck's constant. 
As usual, the quantization map does not generally preserve the
classical magnetic Poisson brackets.
However, it does represent the 
brackets \eqref{eq:xpmagPoisson} of local coordinates as the quantum
commutators 
\bea
[\calo_{x^i},\calo_{x^j}]_\circ=0 \ , \quad
[\calo_{x^i},\calo_{p_j}]_\circ=\ii\hbar\, \delta^i{}_j \qquad \mbox{and}
\qquad [\calo_{p_i},\calo_{p_j}]_\circ=-\ii\hbar\, B_{ij}(\calo_x) \ .
\label{eq:quantlinear}\eea

To study the representation of the quantum kinematics, it is useful to introduce, for any $\hbar\neq0$, the \emph{magnetic
  translation operators}
\bea
\calp_{ v}=\exp_\circ\big(\tfrac\ii\hbar\, \calo_{ p\cdot  v}\big) \ ,
\eea
where the operation $\exp_\circ$ denotes the usual formal power series expansion
of the exponential function with powers taken with the composition
$\circ$. By the commutation relations
\eqref{eq:quantlinear},  they formally implement translations on configuration
space by constant vectors $v\in\real^d$ at the quantum level:
\bea
\calp_{ v}^{-1}\circ \calo_{x^i}\circ\calp_{ v}=\calo_{x^i+v^i} \ ,
\label{eq:calptransl}\eea
where the inverse is taken with respect to $\circ$.
As noticed originally by~\cite{Jackiw1985}, the operators $\calp_v$ do
not represent the translation group $\real^d$, as expected because the
background 2-form field $B$ breaks the translational symmetries of the
dynamical system on $\real^d$. Instead, by formally iterating the conjugation action
\eqref{eq:calptransl} using \eqref{eq:quantlinear}, one finds that they satisfy the composition and
association relations
\begin{align}
\calp_{ w}\circ\calp_{ v}&= {\mit\Pi}_{v,w}(x) \ \calp_{ v+ w} \ ,
                           \nonumber \\[4pt] \calp_{ w}\circ(\calp_{ v}\circ\calp_{ u}) &=
{\mit\Omega}_{u,v,w}(x) \ (\calp_{ w}\circ\calp_{ v})\circ\calp_{ u} \ ,
\label{eq:magtranslrels}\end{align}
where 
\bea
{\mit\Pi}_{v,w}(x) = \exp\Big(-\frac\ii\hbar\, \int_{\triangle^2( x; v,w)}\, B\Big)
\eea
is a phase factor determined by the magnetic flux through the triangle $\triangle^2( x; v,w)$
based at $x$ and spanned by the translation vectors $v,w$, while
\bea
{\mit\Omega}_{u,v,w}(x) = \exp\Big(\frac\ii\hbar\, \int_{\triangle^3(x;u,v,w)}\, H\Big)
\eea
is a phase factor determined by the total magnetic charge enclosed by the tetrahedron
$\triangle^3(x;u,v,w)$ based at $x$ and spanned by the vectors $u,v,w$
in $\real^d$, or equivalently by Stokes' theorem, the total flux
through the faces of the tetrahedron. In general,
${\mit\Pi}_{v,w}(x)$ defines a 2-cochain of the translation group $\real^d$
with coboundary the 3-cocycle ${\mit\Omega}_{u,v,w}(x)$ in the group
cohomology of $\real^d$ with values in $C^\infty(M,U(1))$. From this
point of view, defining a quantization map means providing a
representation, in a suitable sense, of the algebra \eqref{eq:magtranslrels} of magnetic
translation operators.

Let us now explicitly confirm these formal calculations in our specific cases of
interest.

\paragraph{Maxwell theory.}

In the $d$-dimensional generalization of the classical Maxwell theory, the
magnetic field $B$ on $M=\real^d$ is a closed 2-form, $\dd B=0$, so
there exists a globally defined 1-form $A$ such that $B=\dd A$. The
quantum theory can then be described via canonical quantization. The
quantization map in this case takes functions on phase
space $\Mcal$ to operators acting on the quantum Hilbert space $\CCH={\rm L^2}(M)$ of square-integrable functions on configuration space with respect to the usual Lebesgue
measure $\dd x$ on $\real^d$, with the composition operation $\circ$
given by the usual associative composition of linear operators. This constructs the
Schr\"odinger representation of the algebra of magnetic translations
where the position and kinematical momentum operators are represented as
\bea\label{eq:Schrodingerrep}
(\calo_{x^i}\psi)(x)=x^i\,\psi(x) \qquad \mbox{and} \qquad
(\calo_{p_i}\psi)(x)=-\ii\hbar\, \frac\partial{\partial x^i}\psi(x) +
A_i(x) \, \psi(x)
\eea
on wavefunctions
$\psi\in\CCH$. The magnetic translation
operators in this case can be represented by the Wilson lines
\bea\label{eq:Maxwellmagtransl}
(\calp_{ v}\psi)( x)=\exp\Big(-\frac\ii\hbar\, \int_{[x-v,x]}\,
A\Big)\, \psi( x- v) \ ,
\eea
where $[x-v,x]$ is the straight line in $\real^d$ from the point $x-v$
to $x$. An explicit calculation verifies the composition relation in
\eqref{eq:magtranslrels} in this case where, since $H=\dd B=0$, the
coboundary ${\mit\Omega}_{u,v,w}(x)=1$ is trivial. Thus in this instance
${\mit\Pi}_{v,w}(x)$ defines a 2-cocycle on $M=\real^d$ with values in
$C^\infty(M,U(1))$, and the magnetic translation operators generate a
weak projective representation of the translation group $\real^d$ on
$\CCH$, the adjective `weak' refering to the $x$-dependence of the
2-cocycle that modifies the usual cocycle condition by the action of
$\real^d$ on ${\mit\Pi}_{v,w}(x)$ (see e.g.~\cite{Bunk:2018qvk}). In the case where the components of
the 2-form $B$ are constant, the 2-cocycle simplifies to
\bea
{\mit\Pi}_{v,w} = \e^{-\frac\ii{2\hbar}\, B(v,w)}
\eea
independently of $x$, and we obtain the standard projective
representation of translations in the background of a constant
magnetic field.

The magnetic translation operators also provide the general form of
the quantization map from generic phase space functions $f\in
C^\infty(\Mcal)$, which are integrable with respect to the Lebesgue measure $\dd
X$ on $\Mcal$, to trace-class
operators $\calo_f$ on the Hilbert space $\CCH$. This is called the
magnetic Weyl correspondence, and it is determined by a family of
operators ${\mit\Delta}_\calo(X)$ on $\CCH$ parameterized by points $X=(x,p)$ of
$\Mcal$ which are defined on wavefunctions $\psi\in\CCH$ by
\bea\label{eq:Weylops}
\big({\mit\Delta}_\calo(x,p)\psi\big)(y)=\e^{\frac{\ii\hbar}2\,p\cdot x}\,
\e^{-\ii p\cdot y} \, (\calp_x\psi)(y) \ .
\eea
From \eqref{eq:magtranslrels} it follows that this operator satisfies
\bea
{\mit\Delta}_\calo(X)^\dag = {\mit\Delta}_\calo(-X) \ .
\eea
Then the magnetic Weyl correspondence defines operators $\calo_f$ given by
\bea\label{eq:calof}
\calo_f=\int_\Mcal \ \bigg( \int_\Mcal\, \e^{\ii\omega_0(X,Y)}\, f(Y)
\ \frac{\dd Y}{(2\pi)^d}\bigg)\, {\mit\Delta}_\calo(X) \ \frac{\dd X}{(2\pi)^d}
\eea
for suitable class functions $f\in C^\infty(\Mcal)$, where $\omega_0$ is the Darboux 2-form \eqref{eq:Darboux2form}. The
magnetic Weyl correspondence $f\mapsto\calo_f$ is unitary,
\bea
\calo_{f(X)}{}^\dag = \calo_{\overline{f(-X)}} \ , 
\eea
and it is
  invertible as a consequence of completeness of the Hilbert space $\hil$. Its inverse can be used to
pull back the product of two operators $\calo_f,\calo_g$ to a noncommutative
product of the corresponding phase space functions $f,g$. This defines
the associative magnetic Moyal-Weyl star product through
$\calo_{f\star_B g}=\calo_f \,\calo_g$, which when $B$ is constant is
given explicitly by the twisted convolution product
\bea\label{eq:MoyalWeylstar}
(f\star_B g)(X)=\frac1{(\pi\,\hbar)^d}\, \int_\Mcal \ \int_\Mcal \,
\e^{-\frac{2\ii}\hbar\,\omega_B(Y,Z)} \, f(X-Y)\, g(X-Z) \ \dd Y \ \dd
Z \ ,
\eea
where $\omega_B$ is the twisted symplectic 2-form \eqref{eq:omegaB}; see e.g.~\cite{Bunk:2018qvk} for the explicit expression in the
general case.
This generalizes the standard Moyal-Weyl star product on canonical
phase space for $B=0$, while for any 2-form $B$ it recovers the
ordinary convolution product between integrable functions when $\hbar=0$.

This shows that the magnetic translation operators also provide a
bridge between the canonical formulation of quantum mechanics and the somewhat less known
phase space formulation of quantum mechanics, which is historically
the origin of deformation quantization and star products, see
e.g.~\cite{Zachos2001}. In the phase space quantization scheme, the operators
are the functions $f$ themselves, with the composition
operation $\circ$ given by the star product $\star_B$. The deformation from classical
to quantum observables is then 
characterized by the noncommutative and associative magnetic
Moyal-Weyl 
star product of functions on phase space, which substitutes the
pointwise product and contains the necessary information about the
corresponding quantum system. In particular, the star product deforms the
magnetic Poisson brackets in such a way as to contain all information
related to the commutator of quantum operators $\calo_f$.
All of canonical quantum
mechanics can be rephrased in the language of phase space quantum
mechanics, with some caveats, through a dictionary which for later use
we briefly
summarise as follows:
\begin{itemize}
\item Wavefunctions are given by arbitrary complex-valued functions on phase space
  $\Mcal$, while observables are real functions.
\item The noncommutative composition of operators becomes the
  noncommutative star product of functions.
\item Traces of operators become integrals of functions over $\Mcal$.
\item There is a state function ${\mit\Sigma}\geqslant0$ on $\Mcal$ which is
  normalized, $\int_\Mcal\, {\mit\Sigma}(X)\ \dd X=1$, and which plays
  the role of a density matrix determining the correlations of the
  quantum system. The state function originates
  through the inverse of the magnetic Weyl correspondence
  $f\mapsto\calo_f$ in \eqref{eq:calof} taking quantum operators on
  $\CCH$ to functions on phase space, which determines the magnetic
  Wigner functions representing the density matrices that are quasi-probability distributions on
  $\Mcal$. They heuristically give the
  quantization of the constant energy surfaces in phase space through
  the star product.  Expectation values of operators
  $\calo_f$ in the state $\mit\Sigma$ are given by
\bea
\langle \calo_f\rangle_{\mit\Sigma} = \int_\Mcal\, (f\star_B{\mit\Sigma})(X) \ \dd X \ .
\eea
\end{itemize}

\paragraph{Dirac monopoles.}

Quantization in the case when $H=\dd B\neq0$ is not straightforward, because the operator/state formulation of canonical quantum mechanics cannot handle nonassociative magnetic Poisson brackets: Operators acting on a separable Hilbert space always associate. The exception is the case of the Dirac monopole field \eqref{eq:Diracfield}. Recall that in this case the magnetic Poisson brackets are associative away from location of the monopole at the origin, i.e. on $M^\times = \real^3\setminus\{\,\vec0\,\}$. Recall also from the analysis of Section~\ref{sec:classical} that the electric charge in this background follows a classical trajectory that never reaches the monopole. In the quantum theory this means that the wavefunction of the electric charge should vanish at the origin $\vec0$, so that the quantum mechanics is also confined to $M^\times$.

To understand how to quantize the dynamics on $M^\times$, we observe
that the quantization of Maxwell theory discussed above can be recast
in a more geometric way: The globally defined 1-form potential $A$ can
be thought of as a connection on a (necessarily trivial) line
bundle $L=M\times\complex$ on $M$, with field strength the magnetic
field $B=\dd A$. Then the quantum Hilbert space of states can be
thought of as the space $\hil={\rm L}^2(M,L)$ of square-integrable
(global) sections of this line bundle, the Schr\"odinger
representation of the kinematical momentum operators from
\eqref{eq:Schrodingerrep} can be written in terms of covariant
derivatives $D_A$ as $\calo_{p_i}\psi=(D_A)_i\psi$ for $\psi\in\hil$, and the magnetic translation operators \eqref{eq:Maxwellmagtransl} define parallel transport in $L$. This is just a reformulation of the canonical quantization problem as geometric quantization, which is of course superfluous on $M=\real^d$. 

However, this has the virtue of generalizing to non-trivial
configuration spaces such as $M^\times =
\real^3\setminus\{\,\vec0\,\}$, which is topologically a 2-sphere
$S^2$. The 2-form corresponding to the Dirac monopole field can be
expressed from \eqref{eq:Diracpotential} as $B_g=\dd A_{g,\vec n}$,
where the locally defined 1-form $A_{g,\vec n}$ defines a non-trivial connection on a line bundle $L\to M^\times$, and the quantum Hilbert space is again the space of square-integrable (local) sections $\hil={\rm L}^2(M^\times,L)$. Line bundles on a 2-sphere can be non-trivial with a non-zero Chern number $n\in\zed$ that is obtained by integrating the field strength over $S^2$. Integrating the 2-form $B_g$ over $S^2$ gives $2\,e\,g/\hbar$, which is hence the field strength of a line bundle $L$ if and only if
\bea
n=\frac{2\,e\,g}\hbar\ \in\ \zed \ .
\label{eq:Diracquant}\eea
This is the celebrated \emph{Dirac quantization} of electric
charge, formulated geometrically in this way originally
by~\cite{Wu:1976ge}. The quantization condition \eqref{eq:Diracquant}
in turn implies that the 3-cocycle ${\mit\Omega}_{u,v,w}=1$ in
\eqref{eq:magtranslrels} is trivial, and hence that the corresponding
magnetic translation operators, defined by parallel transport in the
non-trivial line bundle $L$, associate, consistently with their
representability on the Hilbert space $\hil$. Turning this argument
around, the associativity of magnetic translations is equivalent to
${\mit\Omega}_{u,v,w}=1$, which implies the Dirac quantization
condition \eqref{eq:Diracquant}, formulated algebraically in this way
originally by~\cite{Jackiw1985} (see also~\cite{Jackiw:2002wf}). The
quantization map $f\mapsto\calo_f$ in this case can be constructed via
a suitable magnetic Weyl correspondence ${\mit\Delta}_\calo(X)$ on the phase space $\Mcal^\times$ of the configuration space $M^\times$, which induces an associative phase space star product constructed explicitly by~\cite{Carinena:2009ug,Soloviev:2017nwk}.

\subsection{Nonassociative quantum mechanics\label{sec:NAQM}}

For generic {smooth distributions} of magnetic charge
$H\in\Omega^3(M)$, standard canonical quantization breaks
down. However, as first observed by~\cite{Mylonas2012}, the magnetic
Poisson brackets in these instances can still be quantized by
appealing to the formalism of \emph{deformation quantization}. For any
smooth 3-form $H=\dd B\in\Omega^3(M)$, Kontsevich formality provides a
noncommutative and nonassociative star product on functions in
$C^\infty(\Mcal)$ given as a formal power series in $\hbar$; we refer
to the original treatment of~\cite{Mylonas2012} and to the
review~\cite{Szabo:2018hhh} for details of the Kontsevich
expansion. For generic $H$, the Kontsevich formula can only be written
as a formal power series expansion and computed order by order in the
deformation parameter $\hbar$, which becomes increasingly complicated
as the order increases. 

When the components of $H$ are
constant, we can choose the magnetic field to have components
\bea
B_{ij}(x) = \tfrac13\, H_{ijk}\, x^k \ ,
\eea
which is the generalization of the rotationally symmetric magnetic
field \eqref{eq:Brot} in three dimensions. In this case
the Kontsevich series can be summed explicitly to give a closed formal expression for
the nonassociative
star product in terms of the bidifferential operator determined by
the magnetic Poisson bivector $\theta_B$ from Section~\ref{sec:magbrackets}~\cite{Mylonas2012}. In~\cite{Mylonas2013} it was shown
that this star product can be written in the form of a twisted
convolution product which gives a convergent expression (as opposed to
an asymptotic series) and reads as
\bea\label{eq:NAstar}
(f\star_H g)(X) = \frac1{(\pi\,\hbar)^d}\, \int_\Mcal \ \int_\Mcal \,
\e^{-\frac{2\ii}\hbar\,\omega_B(Y,Z)} \, f(X-Y)\, g(X-Z) \ \dd Y \ \dd
Z \ ,
\eea
for suitable class functions $f,g\in C^\infty(\Mcal)$. Note that this is
formally identical to the star product \eqref{eq:MoyalWeylstar} which
is written for constant $B$, with the difference here that the
star product \eqref{eq:NAstar} is \emph{nonassociative}, with the
nonassociativity controlled by the constant magnetic charge $H$. This
star product was subsequently computed by a variety of other means,
see e.g.~\cite{BL,Kupriyanov2015}.

In this framework, the quantum operators $\calo_f$ are the functions
$f\in C^\infty(\Mcal)$ themselves, and the composition operation
$\circ$ is the nonassociative star product $\star_H$. The
nonassociative magnetic translation operators in this case are given
by the phase space functions
\bea
\calp_v:=\e^{\frac\ii\hbar\,p\cdot v} \ .
\eea
They satisfy the algebraic relations \eqref{eq:magtranslrels} with
\begin{align}
\calp_w\star_{ H}\calp_v &= {\mit\Pi}_{v,w}(x) \ \calp_{v+w} \ ,
                           \nonumber \\[4pt]
\calp_w\star_{ H}\big(\calp_v \star_{ H} \calp_u\big) &=
{\mit\Omega}_{u,v,w} \ \big(\calp_w \star_{  H} 
\calp_v \big) \star_{ H}\calp_u \ ,
\end{align}
where 
\bea
{\mit\Pi}_{v,w}(x)=\e^{-\frac{\ii}{6\hbar}\, H(x,v,w)} \qquad
\mbox{and} \qquad {\mit\Omega}_{u,v,w} = \e^{\frac{\ii}{6\hbar}\,
  H(u,v,w)}
\eea
define a 3-cocycle of the translation group $\real^d$.

Although there is no standard canonical quantization framework
available in this case, one can now adapt the formalism of phase space quantization outlined
in Section~\ref{sec:quantization} using the nonassociative
star product. In~\cite{Mylonas2013} it is shown that the phase space
formulation of nonassociative quantum mechanics obtained in this way
is physically sensible: It passes all preliminary tests for a sensible
quantum theory, such as reality, positivity and completeness of
eigenstates of physical observables (real functions). It is a
completely \emph{quantitative} framework which gives novel
predictions, such as modified uncertainty
relations. See~\cite{Szabo:2017yxd} for a discussion of some of the
interesting predictions for the physics of electron propagation in
fields of magnetic
monopoles; we shall give a concrete
application later on to explain some of the unusual physics in
locally non-geometric backgrounds of string theory. This
nonassociative quantum theory is in no conflict with the studies of
Jordanian quantum mechanics discussed in Section~\ref{sec:Jordan},
because the algebra of functions on phase space with the
nonassociative star product \eqref{eq:NAstar} is \emph{not} an
alternative algebra: This was shown in~\cite{Bojowald:2016lnl} for a class of
nonassociative star products which includes \eqref{eq:NAstar}, and
subsequently proven to be a general feature of nonassociative
deformation quantization by~\cite{Vassilevich:2018sxx}.
At leading semi-classical order, the nonassociative algebra of
functions is alternative, and this may be used to study a
noncommutative Jordan algebra whose quantum moments can be developed
perturbatively in $H$~\cite{Bojowald:2014oea}. This formalism was
subsequently used to develop certain testable features of
nonassociative quantum mechanics
in~\cite{Bojowald:2015cha,Bojowald:2018qqa}. 
In particular, the
possibility of producing magnetic monopoles in particle colliders has
been considered extensively, see e.g.~\cite{Gould:2019myj} for a recent discussion
and further references. In this way the present treatment of the physics of smooth
distributions of magnetic monopoles may be of use to
develop measurable predictions of nonassociativity, such as in the
MoEDAL experiment at the Large Hadron Collider (LHC) which aims to pursue the quest for
magnetic monopoles and dyons at LHC
energies~\cite{Acharya:2017cio,Baines:2018ltl}.

One undesirable feature of this deformation quantization approach,
aside from the usual technical and conceptual issues arising in
conventional phase space quantum mechanics~\cite{Zachos2001}, is that for magnetic
charges beyond the constant case, the star product can only be
developed as an asymptotic series in Planck's constant $\hbar$. From this point of view deformation quantization is not a true quantization in the
physical sense, where
Planck's constant can be set to some fixed finite value. It would
therefore still be desirable to have an alternative form of canonical quantum
mechanics available, in order to pursue better the foundations of this
nonassociative quantum theory, and to seek further testable aspects of
the model. What is missing in the theory is a nonassociative analog of
the magnetic Weyl correspondence ${\mit\Delta}_\calo(X)$, taking phase space
functions to operators on a ``Hilbert space''. We conclude this section by briefly summarising three such
attempts at a Hilbert space formulation of nonassociative quantum
mechanics.

\paragraph{Symplectic realization.}

Symplectic realization is a well-known method from Poisson geometry
that enables one to set up the problem of quantizing generic Poisson
brackets in the realm of geometric quantization of phase spaces. 
In~\cite{Kupriyanov:2018xji} it was demonstrated how to extend the
technique to the case of magnetic Poisson brackets. The idea is to ``double''
the phase space $\Mcal$ to an
  extended phase space with coordinates $(x^i,\tilde x^i,p_i,\tilde
  p_i)$ and the
  symplectic brackets
\begin{align}
\{x^i,p_j\}&=\{\tilde x^i,p_j\}=\{x^i, \tilde p_j\}=\delta^i{}_j \ ,
             \nonumber \\[4pt]
\{p_i,p_j\}&=B_{ij}(x)+\mbox{$\frac12$}\,\tilde
             x{}^k\,\big(\partial_kB_{ij}(x)-H_{ijk}(x)\big) \ ,
             \nonumber \\[4pt]
\{p_i,\tilde p_j\}&=\{\tilde p_i,p_j\}=\mbox{$\frac12$}\,B_{ij}(x) \ .
\label{eq:symplreal}\end{align}
These brackets define an associative algebra and are inverse to a
symplectic 2-form which pulls back to the twisted 2-form $\omega_B$
under the map which embeds the original phase space $\Mcal$ into the
extended phase space, or more concretely which restricts to $\omega_B$
by eliminating the auxiliary coordinates $(\tilde x^i,\tilde
p_i)$. Some mathematical aspects of this symplectic realization are
reviewed in~\cite{SzaboDurham}.

On the extended phase space one can then introduce
the $O(d,d){\times} O(d,d)$-invariant Hamiltonian
\bea
\widehat{\sf H}=\frac1m\, p_I\,\eta^{IJ}\,p_J \ ,
\eea
where $ p_I=(p_i,\tilde
p_i)$ and
\bea\label{Oddmetric}
\eta=\bigg(\begin{matrix} 0 \ \ & \ 1 \\ 1 \ \ & \ 0  \end{matrix}\bigg)
\eea
is the $O(d,d)$-invariant metric. The corresponding
Hamilton equations of motion for $d=3$ with the brackets \eqref{eq:symplreal} then reproduce the Lorentz force law
\eqref{eq:Lorentzforce} and \eqref{eq:kinmom} for the physical degrees
of freedom $(x^i,p_i)$. This mimics the situation in double field
theory~\cite{HullZwiebach}, with one crucial difference. A consistent Hamiltonian
reduction of the auxiliary degrees of freedom, via the imposition of suitable
first or second class constraints, is possible if and only if there is no
magnetic charge, $H=0$: There is \emph{no} polarisation of the
extended symplectic algebra which is consistent with both the Lorentz
force law and the nonassociative magnetic Poisson
brackets~\cite{Kupriyanov:2018xji}. This feature is seemingly related
to the dissipative nature of the classical dynamics in a rotationally
symmetric magnetic field that we discussed in
Section~\ref{sec:classical}, since a proper Hamiltonian formulation of
dissipative systems also requires the introduction of auxiliary
variables, representing a reservoir, in order to conserve the total
energy. However, in the present case the total energy is already
conserved, so the meaning of this analogy is not entirely clear. Thus
while the formalism of symplectic realisation formally solves the
problem of a Hilbert space formulation of the nonassociative quantum
mechanics, the meaning of the auxiliary degrees of freedom, which
cannot be eliminated, remains unclear. See the contribution of
V.~Kupriyanov to these proceedings for further details of this method.

\paragraph{2-Hilbert spaces.}

Nonassociativity implies the appearence of \emph{higher structures} in
algebra, where in addition to non-vanishing binary commutators of a
multiplication one encounters non-zero ternary Jacobiators (or
associators). These structures are natural from the point of view of
the theory of 
monoidal categories, wherein one works with algebra objects not in the
usual category of vector spaces, where associativity of tensor
products holds on the nose, but in a category where associativity
holds only up to a natural isomorphism called the associator. From the perspective of geometric quantization, this
corresponds to the passage from line bundles on $M$, with 2-form field strength
$B=\dd A$ satisfying the Bianchi identity $\dd B=0$, to \emph{gerbes} on $M$ whose
field strength is controlled by a closed 3-form $H=\dd B\neq0$. In this case
the usual quantum Hilbert space of sections of a line bundle should be
replaced by an analog of sections of a gerbe, that has the structure of
a `categorified' Hilbert space called a 2-Hilbert space which is a certain monoidal category. Geometric
quantization of the magnetic Poisson brackets for any magnetic field
$B$ was worked out in this
setting by~\cite{Bunk:2018qvk}, where the 3-cocycles
${\mit\Omega}_{u,v,w}(x)$ appearing in \eqref{eq:magtranslrels} are
interpreted as controlling suitable higher versions of a (weak) projective
representation, implemented by magnetic translation functors $\calp_v$
on the 2-Hilbert space; this utilizes a higher notion of parallel
transport in a gerbe. While again this approach formally captures a
Hilbert space formulation of the nonassociative quantum mechanics, the
physical meaning and development of the higher structures involved are
at present unclear. A concise review of this approach can be found
in~\cite{SzaboDurham}.

\paragraph{Transgression to loop space.}

In higher geometry it is well-known that the nonassociativity features
of a gerbe on $M$ can be traded for more conventional noncommutative
features of a line bundle on the loop space $C^\infty(S^1,M)$ of
embeddings of a circle
$S^1$ into the configuration space $M$; this mapping is called
transgression and it has a natural
interpretation of trading particle degrees of freedom for closed
string degrees of freedom. The field strength of a gerbe on $M$ is sent to the field
strength of a line bundle on the closed string configuration space $C^\infty(S^1,M)$ under the transgression
map, and on the latter one can proceed to apply the standard
techniques of geometric quantization; this approach was pursued
in~\cite{Saemann2011,Saemann13}. However, this approach simply hides
the higher structure of the 2-Hilbert space of sections of a gerbe
in a Hilbert space of sections of a line bundle over an
infinite-dimensional configuration space of closed strings, again
making explicit constructions and their physical interpretations
unwieldy. Once more, although the problem of a Hilbert space approach
to quantization is formally solved, many technical and conceptual
difficulties remain awaiting for further exploration. The discussion
of this paragraph motivates considerations of these higher
nonassociativity structures in
terms of closed strings, to which we now turn.

\section{Closed strings in locally non-geometric flux backgrounds}
\label{sec:stringsflux}

\subsection{The $R$-flux model\label{sec:Rfluxmodel}}

We now turn to our attention to somewhat more conjectural applications to non-geometric
string theory. At the algebraic level, the pertinent phase space
brackets are obtained by applying a certain`duality' transformation to
the magnetic Poisson brackets from Section~\ref{sec:magbrackets}. For
this, we invoke \emph{Born reciprocity}, which interchanges the roles
of local position and momentum coordinates through the mapping
$(x,p)\mapsto(p,-x)$ on phase space $\Mcal$ of order~$4$. This map
preserves the canonical symplectic 2-form $\omega_0$ on $\Mcal$, but
it does not preserve the twisted 2-form $\omega_B$, and so is not a natural symmetry of the magnetic Poisson brackets. Instead, it has a
natural geometric interpretation by regarding the phase space
$\Mcal=T^*M$ as a para-Hermitian manifold~\cite{Marotta:2018myj} where
it corresponds to an $O(d,d)$-transformation which preserves the
Lorentzian metric \eqref{Oddmetric} and takes the magnetic
Poisson brackets \eqref{eq:xpmagPoisson} of local coordinates,
depending on a 2-form $B\in\Omega^2(M)$ on configuration space $M$, to
the twisted Poisson brackets
\bea
\{x^i,x^j\}_\beta=-\beta^{ij}(p) \ , \qquad
\{x^i,p_j\}_\beta=\delta^i{}_j \qquad \mbox{and} \qquad
\{p_i,p_j\}_\beta=0 \ ,
\eea
depending on a 2-form $\beta\in\Omega^2(M^*)$ on the dual momentum
space $M^*=\real^d$. In this case the twisting is by a 3-form on momentum
space $M^*$, called the `$R$-flux' $\ell_s^3\,R=\dd\beta\in\Omega^3(M^*)$, where $\ell_s$ is the string length. This
now gives a nonassociative configuration space with the nonvanishing coordinate
Jacobiators
\bea\label{eq:RfluxJac}
\{x^i,x^j,x^k\}_\beta = -\ell_s^3\, R^{ijk}(p) \ .
\eea
These brackets have a natural formulation as a 2-term $L_\infty$-algebra which was developed in~\cite{Mylonas2012,Hohm:2017cey}.

The dynamical system captured by these dual brackets is called the
\emph{$R$-flux model}, and it conjecturally describes the phase space
of closed strings propagating in `locally non-geometric' $R$-flux
backgrounds of string theory, which have no formulation as a
conventional spacetime~\cite{stnag2,Mylonas2012}. As for the case of
the magnetic Poisson brackets, the prototypical example arises in
$d=3$ spacetime dimensions with constant $R$-flux. This originates
from a type~II background containing a 3-torus $M=T^3$ with constant $H$-flux (proportional to the volume form) through a chain of successive T-dualities which gives rise to geometric and non-geometric fluxes that are depicted schematically as~\cite{Shelton:2005cf}
\bea
{H_{ijk}} \ \stackrel{{\sf T}_i}{\longleftrightarrow} \ {f^i{}_{jk}} \ \stackrel{{\sf T}_j}{\longleftrightarrow} \ {Q^{ij}{}_k} \ \stackrel{{\sf T}_k}{\longleftrightarrow}\ {R^{ijk}} \ , 
\eea
where ${\sf T}_i$ denotes a T-duality transformation along the $i$-th
coordinate direction which sends string winding numbers $(w^i)\in
H_1(T^3,\zed)=\zed\oplus\zed\oplus\zed$ to momentum modes $(p_i)$ (and
vice-versa). Starting from the geometric background $(T^3,H)$, a
T-duality takes this to the Heisenberg nilmanifold $\tilde T^3$, which is a
geometric background with no $B$-field and a metric flux (torsion)
$f$. Applying another T-duality then results in a
T-fold~\cite{Hull:2004in}, a non-geometric background which is not
globally well-defined  with the transition functions controlled by a
`$Q$-flux'. A final T-duality then results in a background which
cannot be described even locally in conventional geometric terms,
characterised by an $R$-flux. The triple T-duality transformation
${\sf T}_{ijk}$ which sends the $H$-flux 
\bea
{H_{ijk} = \partial_{[i}B_{jk]} }
\eea
to the $R$-flux
\bea
{R^{ijk} =\hat\partial^{[i}\beta^{jk]}}
\eea
is best understood in double field theory~\cite{Andriot:2012an}, where it maps the 2-form NS--NS $B$-field to a bivector $\beta$ with $\hat\partial{}^i$ denoting `modified derivatives' with respect to the coordinates dual to the string winding modes. This is completely analogous to the $O(d,d)$-transformation applied to the magnetic Poisson brackets above.

In the quantum theory, a suitable substitute for canonical quantization of locally non-geometric closed strings is provided by the deformation quantization of Section~\ref{sec:NAQM}, with the replacements of phase space coordinates $X^I=(p_i,-x^i)$ and the magnetic field $B$ with $\beta$ everywhere; in the case of a constant $R$-flux, we denote the corresponding nonassociative star product obtained from \eqref{eq:NAstar} with these substitutions by $\star_R$. The phase space formulation of nonassociative quantum mechanics discussed in Section~\ref{sec:NAQM} can then be used to quantitatively explore the physical implications of the local non-geometry of the configuration space~\cite{Mylonas2013}. As an application, let us compute the expectation values of the oriented volume uncertainty operators 
\bea
V^{ijk}=\big\langle \tfrac12\,[\Delta x^i,\Delta x^j,\Delta
x^k]_{\star_R} \big\rangle_{\mit\Sigma} = \int_{\Mcal}\, \big(\tfrac12\,[\Delta x^i,\Delta x^j,\Delta x^k]_{\star_R}\, \star_R\, {\mit\Sigma}\big)(X) \ \dd X
\eea
determined by the Jacobiators of the position uncertainty operators $\Delta x^i=x^i-\langle x^i\rangle_{\mit\Sigma}$ for an arbitrary state function $\mit\Sigma$. This quantity computes the quantum volume of the tetrahedron in configuration space spanned by the coordinate uncertainties in the given directions. While this quantity would obviously vanish in the associative case, in the present case it can be explicitly computed to give a non-zero quantum of volume~\cite{Mylonas2013}
\bea
V^{ijk}=\frac{\hbar^2 \,\ell_s^3}2 \, R^{ijk} \ ,
\eea
independently of the chosen state $\mit\Sigma$.
This corroborates the expectations that particles cannot be used as
probes of the $R$-flux background~\cite{Wecht:2007wu}. This can be
argued in $d=3$ dimensions via the Freed-Witten anomaly, which forbids
D3-branes from wrapping a 3-torus $T^3$ with nonvanishing
$H$-flux. Under a triple T-duality ${\sf T}_{ijk}$, D3-branes map to D0-branes and the $H$-flux to the $R$-flux, and hence D0-branes cannot propagate in locally non-geometric string backgrounds.

\subsection{How closed strings see nonassociativity\label{sec:triproducts}}

The nonassociative phase space structure of the $R$-flux model raises
some curiosities concerning basic string physics, insofar that at the
level of the two-dimensional worldsheet conformal field theory one
does not expect any such structures to arise (these would otherwise
violate axioms of conformal field theory). The issue was clarified
by~\cite{Aschieri:2015roa} which demonstrated explicitly how to pass
from nonassociative phase space star products to certain ternary
products which capture the nonassociativity of configuration space
fields in a way that is consistent with conformal invariance. For
this, one defines configuration space \emph{triproducts} in the $R$-flux model through
\begin{eqnarray}
(f_1\vartriangle f_2 \vartriangle f_3)(x) = \big(f_1(x)\star_R f_2(x) \big)\star_R f_3(x) \big|_{p=0} 
\end{eqnarray}
for any three functions $f_1,f_2,f_3\in C^\infty(M)$, where a particular bracketing of the nonassociative star product has been chosen. By Fourier transforming the configuration space fields $f_i$ on $M$ to fields $\tilde f_i$ on Fourier space $\tilde M$ one can express this triproduct as a twist of the usual Fourier convolution product of three fields as~\cite{Aschieri:2015roa}
\begin{eqnarray}\label{eq:triproduct}
&& (f_1\vartriangle f_2 \vartriangle f_3)(x) \\ && \qquad = \int_{\tilde M} \ \int_{\tilde M} \ \int_{\tilde M} \, \tilde f_1(k_1)\, \tilde f_2(k_2) \, \tilde f_3(k_3) \, {\e^{-\frac{\ii\hbar^2\,\ell_s^3}{4}\, R(k_1,k_2,k_3)}} \, \e^{\ii(k_1+k_2+k_3)\cdot x} \ \frac{\dd k_1}{(2\pi)^d}\ \frac{\dd k_2}{(2\pi)^d}\ \frac{\dd k_3}{(2\pi)^d}\ .  \nonumber
\end{eqnarray}
One can introduce a natural 3-bracket via antisymmetrization of the triproducts to get
\bea\label{eq:3bracketantisym}
[f_1,f_2,f_3]_{\vartriangle} = \frac13 \, \sum_{\sigma\in S_3}\, (-1)^{|\sigma|}\, f_{\sigma(1)}\vartriangle f_{\sigma(2)}\vartriangle f_{\sigma(3)} \ ,
\eea
where the sum runs through all permutations of degree~$3$. This naturally quantizes the 3-bracket defined by the classical Jacobiator \eqref{eq:RfluxJac} of the $R$-flux model:
\bea
[x^i,x^j,x^k]_{\vartriangle} =\hbar^2 \, \ell_s^3\, R^{ijk} \ .
\eea
In $d=3$ dimensions, the triproduct \eqref{eq:triproduct} was
independently postulated by Takhtajan~\cite{Takhtajan:1993vr} over 20
years earlier as a candidate quantization of the canonical Nambu-Poisson
3-bracket \eqref{eq:NPR3} on $\real^3$.

These triproducts were rediscovered by~\cite{Blumenhagen2011} through
calculations of the scattering of momentum states of closed string tachyon vertex operators
\bea
{V_k(z,\bar z)= \ : \exp\big(\ii k\cdot x(z,\bar z)\big) :} 
\eea
in a linearisation of conformal field theory on flat space in the $R$-flux background. Computing the 3-point functions of such states then reproduces the triproduct to linear order in $R$, and indeed the all orders expression for the multiplication of vertex operators was conjectured to be given by the exact expression \eqref{eq:triproduct}:
\bea
{\big\langle V_{k_1} \, V_{k_2} \, V_{k_3}\big\rangle_R = \exp\Big(-\tfrac{\ii\hbar^2\ell_s^3}{4}\, R(k_1,k_2,k_3)\Big)} \ .
\eea
The precise meaning of this formula was thus subsequently clarified by~\cite{Aschieri:2015roa}: The triproduct \eqref{eq:triproduct} on configuration space is inherited from the nonassociative star product on phase space. Both the star product and the triproduct violate the strong constraint of double field theory~\cite{Blumenhagen:2013zpa,Chatzistavrakidis:2018ztm}, which thus forbids a conventional spacetime interpretation, but on-shell associativity of conformal field theory amplitudes is retrieved as a result of the integral identity~\cite{Aschieri:2015roa}
\bea
\int_M \, (f_1\vartriangle f_2 \vartriangle f_3)(x) \ \dd x= \int_M \, f_1(x)\, f_2(x)\, f_3(x) \ \dd x \ .
\eea
Thus nonassociativity is an \emph{off-shell} effect, invisible to the on-shell closed strings. See~\cite{Szabo:2018hhh} for a detailed discussion of various other caveats associated with these and other derivations of closed string nonassociativity.

\subsection{Nonassociative gravity}
  
Given that the closed string sector contains gravity, it is natural to
ask if there is a consistent nonassociative theory of gravity which
governs the low-energy dynamics of closed strings in locally
non-geometric backgrounds. This endeavour has been pursued at length
by developing a theory of nonassociative (Riemannian) differential
geometry through a generalization of the usual twist deformation
techniques to the case where the twist is provided by a quasi-Hopf
algebra $2$-cochain, whose coboundary is a $3$-cocycle controlling
nonassociativity in a similar way as
previously~\cite{Mylonas2013,Barnes:2014ksa,Aschieri:2015roa,
  Barnes:2015uxa,Barnes:2016cjm, Blumenhagen:2016vpb,
  Aschieri:2017sug}. Progress in this direction has been slow due to
the extremely complicated nature of the technical constructions
required. 

The main result has been the development of a metric
formulation of nonassociative gravity on the phase space of the
$R$-flux model. One can construct a Ricci tensor as well as a unique
metric-compatible torsion-free connection, playing the role of the
Levi-Civita connection in ordinary Riemannian geometry~\cite{Aschieri:2017sug}. Using a
projection to configuration space analogous to that described in
Section~\ref{sec:triproducts}, this results in a non-trivial real
deformation of the spacetime Ricci tensor given by~\cite{Aschieri:2017sug}
\begin{align}
{\rm Ric}^\circ_{ i j} &= {\rm Ric}_{ i j} +
\frac{\hbar^2\,\ell_s^3}{4} \, R^{ a b c}\,\Big(
\partial_ k\big(\partial_ a  g^{ k l}\, (\partial_ b
  g_{ l m})\, \partial_ c {\mit\Gamma}_{ i j}^{m} \big) 
- \partial_ j\big(\partial_ a  g^{ k l}\, (\partial_ b
  g_{ l m})\, \partial_ c {\mit\Gamma}_{ i k}^{  m}\big)
\\[4pt] & \qquad \qquad \qquad \qquad\qquad +\,  \partial_ c   g_{ m n}\big( \partial_ a
(  g^{ l m}\, {\mit\Gamma}_{ l j}^{ k})\, \partial_ b
{\mit\Gamma}_{ i k}^{ n} - \partial_ a
(  g^{ l m}\, {\mit\Gamma}_{ l k}^{ k})\,\partial_ b
{\mit\Gamma}_{ i j}^{ n}
\nonumber \\[4pt] & \qquad\qquad \qquad \qquad \qquad+\, ({\mit\Gamma}_{ i k}^{ l}\, \partial_ a
  g^{ k m} - \partial_ a {\mit\Gamma}_{ i k}^{ l} \,  g^{ k m})\,           
\partial_ b {\mit\Gamma}_{ l j}^{n} - ({\mit\Gamma}_{ i j}^{ l}\, \partial_ a
  g^{ k m} -\partial_a {\mit\Gamma}_{ i j}^{ l}\,  g^{ km})\,              
\partial_b {\mit\Gamma}_{l k}^{n} \big) \Big) \ , \nonumber
\end{align}
where ${\rm Ric}_{ij}$ is the Ricci tensor of a metric $g_{ij}$ on $M$
with Christoffel symbols ${\mit\Gamma}_{ij}^k$. This expression holds to
linear order in the $R$-flux, which is also the order at which the
conformal field theory calculations of~\cite{Blumenhagen2011} are
reliable. The fact that this quantity is real is a remarkable feature
of the theory, as the Ricci tensor on phase space is complex-valued
but its imaginary parts vanish under the projection to spacetime.

The final goal of nonassociative gravity is to mimic what happens in
the well-known case of the open string sector with constant
$B$-field. In that case the low-energy effective theory of D-branes
wrapped on a $d$-dimensional torus $T^d$ is a supersymmetric noncommutative Yang-Mills
theory which is invariant under open string $SO(d,d;\zed)$
T-duality. One then wishes to find an equivalent $O(d,d)$-invariant
(off-shell) nonassociative version of the closed string effective action
\bea
S=\frac1{16\,\pi\,G}\, \int_M\, \sqrt{g}\ \Big({\rm Ric}-\frac1{12}\,
\e^{-\phi/3}\, H_{ijk}\,
H^{ijk}-\frac16\,\partial_i\phi\,\partial^i\phi+\cdots\Big) \ \dd x \ ,
\eea
where $G$ is the gravitational constant, $\phi$ is the string dilaton
field, and we have written only the bosonic part of the full type~II
supergravity action for brevity. Although much remains to be seen from
such an effective theory, we can move on and seek the generalizations
of these string theory backgrounds in their uplifts to M-theory.

\section{M2-branes in locally non-geometric flux backgrounds}
\label{sec:M2flux}

\subsection{M-theory lift of the $R$-flux model}

Type~IIA string theory on a background containing the spacetime $M$
lifts to M-theory on the total space $M'$ of a circle bundle
\bea
\xymatrix{
S^1 \ \ar@{^{(}->}[r] & \ M' \ar[d] \\
 & \ M
}
\eea
where the radius $\lambda$ of the circle fibres geometrizes the string
coupling $g_s$. The lifts of the three-dimensional constant $R$-flux
backgrounds to locally non-geometric backgrounds of M-theory was
considered for compactifications to four dimensions in~\cite{Gunaydin:2016axc} and to
higher dimensions in~\cite{Lust:2017bwq}; we focus here on the former lift. One starts from a double
T-duality taking the Heisenberg nilmanifold $M=\tilde T^3$ to the
$R$-flux background: 
\bea
{f^i{}_{jk}} \ \xrightarrow{ \ {\sf T}_{jk} \ } \
{R^{ijk}=R\,\varepsilon^{ijk}} \ .
\eea
For simplicity we consider the lift of this chain to M-theory on the
trivial circle bundle $M' =M\times S^1$. We
denote the local coordinates on $M$ by $x^i$, $i=1,2,3$, as previously, and on the
M-theory circle $S^1$ by $x^4$; we collect these coordinates into the
four-dimensional coordinate vector $\vec x=(x^\mu)=(x^i,x^4)$ of
$M'$. In this case the T-duality ${\sf T}_{jk}$ lifts to a
U-duality ${\sf U}_{\mu\nu\rho}$ 
which sends membrane wrapping modes $(w^{ij})\in H_2(M',\zed)$ to
  momentum modes $(p_i)$ (and vice-versa). The lift of the
  non-geometric string theory $R$-flux to M-theory can be described
  explicitly in $SL(5)$ exceptional field
  theory~\cite{Blair:2014zba,Chatzistavrakidis:2019seu}. The U-duality
  ${\sf U}_{\mu\nu\rho}$ takes the 3-form $C$-field $C_{\mu\nu\rho}$
  of 11-dimensional supergravity to a trivector
  ${\mit\Omega}^{\mu\nu\rho}$ which is a potential for the M-theory $R$-flux
 \bea
R^{\mu,\nu\rho\alpha\beta}=
\hat\partial^{\mu[\nu}{\mit\Omega}^{\rho\alpha\beta]} \ , 
\eea
  where $\hat\partial{}^{\mu\nu}$ denotes a `modified derivative' with
  respect to the coordinates conjugate to M2-brane wrapping modes;
  this is \emph{not} a $5$-vector but rather a mixed symmetry tensor
  whose first index transforms as a vector under four-dimensional
  rotations in $SO(4)$ and whose remaining four indices transform in
  the totally antisymmetric representation of $SO(4)$. The explicit choice of non-vanishing components
\bea
R^{4,\mu\nu\alpha\beta} = R\,
    \varepsilon^{\mu\nu\alpha\beta} \ ,
\eea
    which we shall often make in
    what follows, then breaks the $SL(5)$ symmetry to $SO(4)$.

The quasi-Poisson brackets of the M2-brane phase space in the
locally non-geometric $R$-flux background in this case were
conjecturally described in~\cite{Gunaydin:2016axc}. Recall from
Section~\ref{sec:Rfluxmodel} that there are no D0-branes permitted on the string
background $M$ with $R$-flux. Since D0-brane charges in type~IIA
string theory lift to momentum modes along the M-theory circle, this
implies that the momentum $p_4=0$ vanishes along the M-theory
direction. We can write this constraint on the locally non-geometric
M-theory background more covariantly in the form
\bea
R^{\mu,\nu\rho\alpha\beta}\, p_\mu=0 \ ,
\eea
which is trivially satisfied in the absence of M-theory $R$-flux, but
otherwise cuts out a one-dim{-}ensional subspace of the phase space. In
this case the membrane is purported to have a seven-dimensional phase
space $\Mcal'$ with quasi-Poisson brackets given by
\begin{align}\label{eq:M2brackets}
\{x^i,x^j\}_{\lambda} &=\ell_s^3\,
                        R^{4,ijk4}\, p_k \qquad \mbox{and} \qquad
                        \{x^4,x^i\}_{\lambda} = \lambda\, \ell_s^3 \,
                       R^{4,1234}\, p^i \nonumber \ , \\[4pt]
{} \{x^i,p_j\}_{\lambda} &= \delta^i{}_j\,x^4+ \lambda\,
                        \varepsilon^i{}_{jk}\, x^k 
                        \qquad \mbox{and} \qquad \{x^4,p_i\}_{\lambda} =
                          \lambda^2\,x_i \ , \nonumber \\[4pt]
{} {\{p_i,p_j\}_{\lambda}} &= {- \lambda\, \varepsilon_{ijk}\, p^k} \ , 
\end{align}
which have the Jacobiators
\begin{align}
{ {} \{x^i,x^j,x^k\}_{\lambda}}&= \ell_s^3\,
                            R^{4,ijk4} \, x^4 \qquad \mbox{and} \qquad 
{} \{x^i,x^j,x^4\}_{\lambda} = -\lambda^2\, \ell_s^3 \, R^{4,ijk4} \,
                                x_k \ , \nonumber \\[4pt]
{ {} \{p_i,x^j,x^k\}_{\lambda} }&= \lambda\,\ell_s^3 \, R^{4,1234}\,
                                 \big(\delta_i{}^j\, p^k-\delta_i{}^k
                                 \, p^j \big) \qquad \mbox{and} \qquad
{} \{p_i,x^j,x^4\}_{\lambda} = \lambda^2\, \ell_s^3 \, R^{4,ijk4}\, p_k
  \ , \nonumber \\[4pt]
{ {} \{p_i,p_j,x^k\}_{\lambda} } &= {-\lambda^2\,
                            \varepsilon_{ij}{}^{k}\, x^4- \lambda\,
                                  \big(\delta_j{}^k\,
                                  x_i-\delta_i{}^k\, x_j \big) \qquad
                                  \mbox{and} \qquad
{} \{p_i,p_j,x^3\}_{\lambda} =  \lambda^3\, \varepsilon_{ijk}\, x^k} \
                                  , \nonumber \\[4pt]
{ {} \{p_i,p_j,p_k\}_{\lambda} } &= { 0 } \ .
\end{align}
These brackets also admit a natural description as an
$L_\infty$-algebra which is discussed in~\cite{Hohm:2017cey}. 
Note that the brackets involving solely the momentum coordinates $p_i$
correspond to a quaternion subalgebra. 

In fact, these brackets originate from the nonassociative alternative
algebra of octonions $\mathbb{O}$ which we discussed in
Section~\ref{sec:octonions} through the redefinition of the imaginary unit octonions given by
\begin{eqnarray}\label{eq:vecxA}
\vec X = (X^A) =(x^i,x^4,p_i)={\mit\Lambda}\, (e_A)  = \frac1{2}\, \Big(\sqrt{\lambda\,
      \ell_s^{3}\, R/3}\ f_i \,,\,
\sqrt{\lambda^3\, \ell_s^{3}\, R/3}\ e_7\,,\, -\lambda\, e_i
\Big) \ .
\end{eqnarray}
The crucial observation of~\cite{Gunaydin:2016axc} is that at $\lambda=0$ these
brackets reduce to those of the magnetic Poisson brackets for the six-dimensional phase space of closed
strings in the $R$-flux background (with the M-theory
circle coordinate
  $x^4=1$ becoming a central element of the algebra). Since the
  contraction parameter $\lambda$ corresponds to the radius of the
  M-theory circle, this limit is the weak coupling limit $g_s\to0$ which reduces M-theory to perturbative type~IIA string theory in the
  $R$-flux background.

\subsection{Quantization of the M2-brane phase space}

The quantization of the quasi-Poisson brackets \eqref{eq:M2brackets} of the M2-brane phase space was carried out in~\cite{Kupriyanov:2017oob}. It is based on the notion of a \emph{$G_2$-structure} which stems from the automorphism group of the algebra of octonions $\mathbb{O}$. Given the octonionic structure constants $\eta_{ABC}$ from \eqref{eq:octonioncomm}, we equip the seven-dimensional real vector space $\real^7$ with the \emph{cross product}
\bea\label{eq:crossproduct}
(\vec K\,\mbf\times_\eta\,\vec
     P\, )_A = \eta_{ABC}\, K_B\, P_C \ , 
\eea
for vectors $\vec K=(K_A)$ and $\vec P=(P_A)$ in $\real^7$. 
This extends the usual cross product $\mbf\times$ on $\real^3$ defined
analogously in terms of the quaternionic structure constants
$\varepsilon_{ijk}$, whose properties can be abstracted into a general
set of axioms independent of dimension~\cite{Salamon:2010cs,Kupriyanov:2017oob}; similarly to the pattern followed by the real normed division algebras, which only exist in dimensions $1$, $2$, $4$ and $8$, cross products only exist on vector spaces of real dimensions $0$, $1$, $3$ and $7$ (where in the first two instances the cross product of two vectors is defined to be $0$). The cross product \eqref{eq:crossproduct} is invariant under the subgroup $G_2\subset SO(7)$ of the rotation group of $\real^7$. A $G_2$-structure on a seven-dimensional real vector space is a cross product that can be brought to the form \eqref{eq:crossproduct} by a suitable change of basis.

The cross product has a natural representation on the octonion algebra $\mathbb{O}$. 
Given a vector $\vec K\in\real^7$, define the octonion element
\bea\label{eq:Xveck}
O_{\vec K}=K^A\,
  e_A \ .
\eea
Then the cross product \eqref{eq:crossproduct} naturally appears as the closure condition of these elements under the octonionic commutator:
\bea
O_{\vec K\,\mbf\times_\eta\,\vec P} = \tfrac12 \, \big[O_{\vec
  K},O_{\vec P} \big] \ .
\eea
Using the alternativity property $(O\,O)\,O=O\,(O\,O)$ for all $O\in\mathbb{O}$, we can unambiguously define powers of any octonion, and via formal power series expansion an octonionic
  exponential $\exp(O)$. For the elements \eqref{eq:Xveck}, one may
  use the unit octonion relations \eqref{eq:octonioncomm} to explicitly compute
\bea\label{eq:octexp}
\exp\big(O_{\vec K}\big)=\cos|\vec K| + \frac{\sin|\vec
  K|}{|\vec K|} \ O_{\vec K} \ ,
\eea
by a calculation completely analogous to the familiar one in quantum mechanics which uses the quaternion relations of the Pauli spin matrices. From \eqref{eq:octexp} one can now compute explicitly the octonionic Baker-Campbell-Hausdorff formula
\bea
\exp\big(O_{\vec K}\big) \, \exp\big(O_{\vec P}\big) = \exp\big(
  O_{\vec\BBB_\eta(\vec K,\vec P\, )}\big) \ ,
\eea
where the expression for the composition element $\vec\BBB_\eta(\vec
K,\vec P)$ is somewhat complicated, see~\cite{Kupriyanov:2017oob} for
its explicit form. With it one can construct an \emph{octonionic Weyl
  correspondence} ${\mit\Delta}_\calo(\vec X)$ between functions on the seven-dimensional M2-brane phase space and octonions, and by pulling back the octonionic products in the usual way one derives a noncommutative and nonassociative star product of functions $f,g\in C^\infty(\Mcal')$ defined via a twisted Fourier convolution product~\cite{Kupriyanov:2017oob}
\bea
(f\star_\lambda g)(\vec X\, ) = \int_{\tilde\Mcal'} \ \int_{\tilde\Mcal'} \, \tilde{f}(
\vec K\, )\, \tilde{g}( \vec P\, )\, \e^{\ii \hbar\, {\vec{\mathscr{
        B}}_\eta({\mit\Lambda}\,\vec K,{\mit\Lambda}\, \vec P\,
    )}\cdot {\mit\Lambda}^{-1}\,\vec X} \ \frac{\dd \vec K}{(2\pi)^7} \ \frac{\dd \vec P}{(2\pi)^7} \ ,
\eea
where the nondegenerate $7\times7$ transformation matrix $\mit\Lambda$ is defined in \eqref{eq:vecxA}. This star product has the non-trivial feature that in the contraction limit it suitably reduces to the star product of the string $R$-flux model~\cite{Kupriyanov:2017oob}:
\bea
\lim_{\lambda\to0} \ (f\star_\lambda g)(\vec X\,) = (f\star_Rg)(X) \ ,
\eea
after setting the central element $x^4=1$.

To understand the spacetime significance of this nonassociative phase space star product, let us introduce the M2-brane triproducts via projection to the four-dimensional configuration space similarly to before:
\bea
\big(f_1\vartriangle_\lambda f_2\vartriangle_\lambda f_3\big)(
\vec x\, ) = \big((f_1\star_\lambda f_2)\star_\lambda f_3
\big)(x^\mu,p_i)\big|_{ p= 0}
\eea
for any three functions $f_1,f_2,f_3\in C^\infty(M')$. Again the explicit form is much more cumbersome and complicated compared to the string theory triproducts, see~\cite{Kupriyanov:2017oob} for the technical details. Defining 3-brackets via antisymmetrization of these triproducts as in \eqref{eq:3bracketantisym}, for $\lambda=1$ one finds that the 3-brackets of local configuration space coordinates are given by
\bea\label{eq:A4}
[x^\mu,x^\nu,x^\alpha]_{\vartriangle_1} = \hbar^2\, \ell_s^3 \,R
\,\varepsilon^{\mu\nu\alpha\beta}\, x^\beta \ ,
\eea
which we recognise as a quantization of the defining relations (up to rescaling) of the 3-Lie algebra $A_4$ which appears as the nonassociative gauge symmetry underlying the theory of a pair of M2-branes in the Bagger-Lambert theory~\cite{Bagger2012}. The classical version of \eqref{eq:A4} defines, after restriction to vectors of unit length $|\vec x\,|=1$, the canonical $SO(4)$-invariant Nambu-Poisson 3-bracket on the 3-sphere $S^3$~\cite{DeBellis:2010pf}. Thus the triproducts in this case provide a definition of a quantum 3-sphere, which as discussed in Section~\ref{sec:NAMtheory} is an important missing ingredient in the understanding of the dynamics of M5-branes in M-theory. These triproducts are naturally inherited from the nonassociative star product on the seven-dimensional M2-brane phase space in the locally non-geometric background.

\section{M-waves in non-geometric Kaluza-Klein monopole backgrounds}
\label{sec:MWflux}

\subsection{Magnetic monopoles in quantum gravity}

Let us recap where we have gotten to in the present exposition. In Section~\ref{sec:magcharge} we considered the magnetic Poisson brackets describing the phase space kinematics of electrically charged particles in magnetic monopole backgrounds. In Section~\ref{sec:stringsflux} we then applied Born reciprocity to map these brackets onto those describing the phase space of closed strings in $R$-flux backgrounds, and in Section~\ref{sec:M2flux} we lifted these brackets to the phase space of M2-branes. In this final section we ask what is the fate of the M2-brane phase space kinematics when we reapply Born reciprocity.

For this, we apply again an order~$4$ transformation
$(x^\mu,p_i)\mapsto (p_\mu,-x^i)$ on $\Mcal'$, together with the substitution of
$\ell_s^3\,R$ with $e\,\rho$, to the brackets \eqref{eq:M2brackets} to get
\begin{align}\label{eq:rhobrackets}
\{x^i,x^j\}_\rho &= \lambda\,
                                 \varepsilon^{ijk}\, x^k \ , \nonumber \\[4pt]
\{p_i,p_j\}_\rho &= -e\,\rho \, \varepsilon_{ijk}\, x^k  \qquad \mbox{and} \qquad
                        \{p_4,p_i\}_\rho = -\lambda\, e\,\rho\, x_i \ , \nonumber \\[4pt]
\{x_i,p_j\}_\rho &= -\delta_{ij}\, p_4- \lambda\,
                        \varepsilon_{ijk}\, p_k \qquad \mbox{and}
                        \qquad \{x^i,p_4\}_\rho = \lambda^2\,x^i \ .
\end{align}
When $\lambda=0$, these reduce to the magnetic Poisson brackets \eqref{eq:xpmagPoisson} for an
electric charge $e$ in a uniform distribution $\rho$ of magnetic
charge on $\real^3$ with the rotationally symmetric magnetic field \eqref{eq:Brot}. For generic values of $\lambda$, the brackets
\eqref{eq:rhobrackets} describe the kinematics of a seven-dimensional
phase space with an ``extra'' momentum mode $p_4$. We would now like
to understand what physics this system represents.

Setting $\rho=0$ in \eqref{eq:rhobrackets} reveals that the brackets
describe a noncommutative but associative deformation of spacetime
whose coordinates $x^i$ generate a quaternion subalgebra with
commuting momenta
\bea
\{p_i,p_j\}_0=\{p_i,p_4\}_0=0 \ .
\eea
The quantity $\lambda^2\,\vec
  p\,^2+p_4^2$ is easily computed to be a central element of the
  algebra \eqref{eq:rhobrackets} for $\rho=0$, so without loss of generality we may
restrict the four-dimensional momentum space to the surface of the
ellipsoid
\bea\label{eq:ellipsoid}
\lambda^2\,\vec
  p\,^2+p_4^2=1 \ .
\eea
Since the momentum $p_4$ commutes with the other momenta, we can solve
this equation for $p_4$ on the upper hemisphere of the ellipsoid and
reduce the remaining brackets of \eqref{eq:rhobrackets} to
\bea
\{x^i,x^j\}_0 = \lambda\, \varepsilon^{ijk}\,
x^k \qquad \mbox{and} \qquad \{x_i,p_j\}_0 =
-
\sqrt{1-\lambda^2\, \vec p\,^2} \ \delta_{ij} -
\lambda\, \varepsilon_{ijk}\, p_k \ .
\eea
For $\lambda=0$ these are just the canonical phase space Poisson
brackets of the configuration space $\real^3$. For $\lambda\neq0$, they are the natural invariant Poisson brackets on the six-dimensional phase space whose momentum coordinates live on the ellipsoid \eqref{eq:ellipsoid}, with the conjugate coordinates $x^i$ acting as covariant derivatives. The quantization of these brackets has appeared before in the context
of a Ponzano-Regge spin foam model of three-dimensional quantum
gravity~\cite{Freidel:2005me}, provided the parameter $\lambda$ is
identified with the Planck length $\ell_{\rm P}$ of the
three-dimensional spacetime through
\bea\label{eq:Planck}
\lambda = \frac{\ell_{\rm P}}{\hbar} \ . 
\eea

Combining these observations together, we may infer that the
uncontracted octonion algebra, after Born reciprocity, is in some
sense related to the dynamics of magnetic monopoles in the spacetime of
three-dimensional quantum gravity with the identification
\eqref{eq:Planck}. The precise gravitational system which realises
this situation was recognised by~\cite{Lust:2017bgx} in the setting of locally
non-geometric M-theory, and we shall now turn to some of the details.

\subsection{The M-wave phase space}

A magnetic monopole can be embedded into type~IIA string theory as a
D6-brane, with the electric probes provided by D0-branes inside the
D6-brane. The D6-brane lifts to M-theory as a Kaluza-Klein monopole,
which is a background of 11-dimensional supergravity given by the metric
\bea\label{eq:KKmetric}
\dd s_{11}^2=\dd s_7^2+U\,\dd\vec x\cdot\dd\vec x+U^{-1}\,\big(\dd
x^4+\vec A\cdot\dd\vec x\big)^2 \ ,
\eea
where $\dd s_7^2$ is the metric of the transverse space to the
M-theory compactification to four dimensions. The three-dimensional
vector potential $\vec A(\vec x)$ and the harmonic function $U(\vec
x)$ are related through
\bea
\vec\nabla\,\mbf\times\,\vec A = \vec\nabla U \qquad \mbox{and} \qquad
\vec\nabla^2U=\rho \ ,
\eea
for a given distribution of magnetic charge $\rho$. The electric probes of this background are M-waves along the M-theory circle $x^4\in S^1$, which lift the D0-branes.

The standard Kaluza-Klein monopole solution corresponds to the embedding of a single Dirac monopole, with singular distribution $\rho(\vec x)=4\pi\,g\,\delta^{(3)}(\vec
x)$, which lifts a single D6-brane. In that case the vector potential $\vec A=\vec A_{g,\vec n}$ is given locally by \eqref{eq:Diracpotential} and the harmonic function $U(\vec x)$ is given by the Green's function of the three-dimensional Laplacian. Then the four-dimensional part of the metric \eqref{eq:KKmetric} is just the standard Taub-NUT metric.

To understand how the purported locally non-geometric background fits into this solution, we first observe that the parameters of type~IIA string theory, the string coupling $g_s$ and the string length $\ell_s$, are related to the parameters of M-theory, the 11-dimensional Planck length $\ell_{\rm P}$ and the radius of the M-theory circle $R_{11}$, through the identifications~\cite{Witten:1995ex}
\bea
\ell_s^2 =\frac{\ell_{\rm
      P}^3}{R_{11}} \qquad \mbox{and} \qquad g_s =\bigg(\frac{R_{11}}{\ell_{\rm
        P}}\bigg)^{3/2} \ .
\eea
The proper reduction which takes M-theory to type~IIA string theory is the limit where $g_s\to0$ and $R_{11}\to0$ with $\ell_s$ finite. This is compatible with the identification \eqref{eq:Planck} of the algebraic contraction parameter $\lambda$ as the Planck length, since then the limit corresponds to $\lambda\sim\ell_{\rm P}\sim
  R_{11}^{1/3}\rightarrow 0$, as required.
  
In the present case we are interested in a uniform distribution of magnetic charge $\rho$, which can be understood as a smearing of Dirac monopoles throughout three-dimensional space. In this instance there is no local expression for the vector potential $\vec A$ or the metric \eqref{eq:KKmetric}, see~\cite{Lust:2017bgx} for a formal non-local expression. This smeared solution is called a \emph{non-geometric Kaluza-Klein monopole}, which lifts a uniform distribution of D6-branes in type~IIA string theory. Geometrically, it corresponds to the passage from the four-dimensional Taub-NUT space, realised as the total space of an $S^1$-bundle over $\real^3$, to the total space of an $S^1$-gerbe over $\real^3$ which realizes the constant magnetic charge~\cite{Lust:2017bgx}, similarly to our discusson of geometric quantization in Section~\ref{sec:NAQM}. This corroborates the interpretation of the brackets \eqref{eq:rhobrackets} as governing the phase space dynamics of the M-waves; that the phase space is now lacking a position coordinate stems from the fact that the wave has no well-defined localised position $x^4\in S^1$.
  
\subsection{The covariant M-theory phase space 3-algebra}

In~\cite{Kupriyanov:2017oob} it was proposed to view the full eight-dimensional phase space $\hat\Mcal'$ of the M-theory compactification to four dimensions, with or without $R$-flux or Kaluza-Klein monopole charge, in terms of a natural 3-algebra structure, akin to those expected to arise in M-theory that we discussed in Section~\ref{sec:NAMtheory}. In~\cite{Lust:2017bgx} it was then subsequently shown how this 3-algebra provides a unified description of the M2-brane and M-wave phase spaces. The starting point is the notion of a \emph{$Spin(7)$-structure} which is defined by collecting all eight generators of the octonion algebra $\oct$ into $\xi_{\hat A}=(\xi_0,\vec\xi\ ) =(1,e_A)$ and defining the 3-brackets
\bea\label{eq:oct3brackets}
\{\xi_{\hat A},\xi_{\hat B}, \xi_{\hat C}\}_\phi =
  \phi_{\hat A\hat B\hat C\hat D}\, \xi_{\hat D} \ ,
\eea
where the self-dual 4-form $\phi$ has components
\bea
\phi_{0ABC} = \eta_{ABC} \qquad \mbox{and} \qquad \phi_{ABCD} = \eta_{ABCD}
\eea
given by the structure constants $\eta_{ABC}$ of the octonionic commutators \eqref{eq:octonioncomm} and the structure constants $\eta_{ABCD}$ of the octonionic Jacobiators \eqref{eq:octonionjac}. The 3-brackets \eqref{eq:oct3brackets} are symmetric under the subgroup $Spin(7)\subset SO(8)$ of the eight-dimensional rotation group and define a ternary operation called a \emph{triple cross product} on the real vector space $\real^8$~~\cite{Salamon:2010cs,Kupriyanov:2017oob}. A $Spin(7)$-structure on an eight-dimensional real vector space is a triple cross product that can be brought to the form \eqref{eq:oct3brackets} by a suitable change of basis.

Let us now map the 3-brackets \eqref{eq:oct3brackets} onto 3-brackets for the eight-dimensional phase space coordinates through the redefinition of the generators of $\oct$ given by
\bea
X^{\hat A}=\big(x^\mu,p_\mu\big)= \big({\mit\Lambda}\,
\vec\xi\,,\,-\tfrac\lambda2\, \xi_0\big) \ ,
\eea
where the invertible $7\times7$ transformation matrix $\mit\Lambda$ is defined in \eqref{eq:vecxA}. This splitting breaks the $Spin(7)$-symmetry to $SO(4)\times SO(4)$ and results in the 3-brackets
\begin{align}
{} \{x^i,x^j,x^k\}_\phi &= -\frac{\ell_s^3}{2} \,
R^{4,ijk4}\, x^4 \qquad \mbox{and} \qquad \{x^i,x^j,x^4\}_\phi  =
\frac{\lambda^2\,\ell_s^3}{2}\, R^{4,ijk4}\, x_k \ , \label{eq:xxx}
\\[4pt] {} \{p^i,x^j,x^k\}_\phi &=
-\frac{\lambda^2\,\ell_s^3}{2}\, R^{4,ijk4}\,
p_4-\frac{\lambda\,\ell_s^3}{2}\, R^{4,ijk4}\, p_k \ ,
\\[4pt]
{} \{p_i,x^j,x^4\}_\phi &=
 -\frac{\lambda^2\,\ell_s^3}{2}\, R^{4,1234}\,
\delta_i^j\, p_4-\frac{\lambda^2\,\ell_s^3}{2}\,
R^{4,ijk4}\, p_k \ , \\[4pt]
 {} \{p_i,p_j,x^k\}_\phi &= \frac{\lambda^2}2\,
\varepsilon_{ij}{}^k\, x^4-\frac{\lambda}2\,\big(\delta_j^k\,
x_i-\delta_i^k\, x_j\big) \ , \label{eq:ppx} \\[4pt]
{} \{p_i,p_j,x^4\}_\phi &= \frac{\lambda^3}2\,
\varepsilon_{ijk}\, x^k \qquad \mbox{and} \qquad
{} \{p_i,p_j,p_k\}_\phi  = 2\,\lambda \,
\varepsilon_{ijk}\, p_4 \ , \label{eq:ppx4} \\[4pt]
{} \{p_4,x^i,x^j\}_\phi &= \frac{\lambda\, \ell_s^3}{2}\,
R^{4,ijk4}\, p_k \qquad \mbox{and} \qquad {} \{p_4,x^i,x^4\}_\phi
=-\frac{\lambda^2\, \ell_s^3}{2}\, R^{4,1234}\, p^i \ ,
\\[4pt]
 {} \{p_4,p_i,x^j\}_\phi &= \frac{\lambda}2\, \delta_i^j\,
x^4+ \frac{\lambda^2}2 \, \varepsilon_i{}^{jk}\, x_k \ , \label{eq:p4px} 
\\[4pt]
{} \{p_4,p_i,x^4\}_\phi &= \frac{\lambda^3}2 \, x_i
                          \qquad \mbox{and} \qquad
{} \{p_4,p_i,p_j\}_\phi =
\frac{\lambda^2}2 \, \varepsilon_{ijk}\, p^k \ . \label{eq:p4px4}
\end{align}
The first set of 3-brackets in \eqref{eq:xxx} for $\lambda=1$ are just the brackets of the 3-Lie algebra $A_4$ (up to rescaling) that we encountered in \eqref{eq:A4}, which reduce to the Jacobiators \eqref{eq:RfluxJac} of the string $R$-flux model when $x^4=1$ is central. These 3-brackets even describe the free eight-dimensional M-theory phase space in the absence of $R$-flux or Kaluza-Klein monopole charge, in which case only the 3-brackets \eqref{eq:ppx}, \eqref{eq:ppx4}, \eqref{eq:p4px} and \eqref{eq:p4px4} are non-zero. These 3-brackets do \emph{not} define a Nambu-Poisson bracket, because the fundamental identity \eqref{eq:fundid} does not hold and its failure is controlled by a higher 5-bracket~\cite{Kupriyanov:2017oob,Lust:2017bgx}. They all vanish in the limit $\lambda\sim\ell_{\rm P}\to0$ where quantum gravitational effects are turned off.

Consider now any constraint
\bea
G\big(X^{\hat 1},\dots,X^{\hat 8}\big) = 0
\eea
on the eight-dimensional phase space $\hat\Mcal'$. It naturally defines a 2-bracket
\bea\label{eq:fgG}
\{f,g\}_G := \{f,g,G\}_\phi
\eea
on functions $f,g\in C^\infty(\hat\Mcal')$, which describes the reduction of $\hat\Mcal'$ with its 3-algebra to a seven-dimensional submanifold with quasi-Poisson brackets \eqref{eq:fgG}. This breaks the $Spin(7)$-symmetry to its $G_2$ subgroup, under which the spinor representation of $Spin(7)$ decomposes according to the branching rule
\bea
\mbf 8\big|_{G_2} = \mbf 7\oplus\mbf 1
\eea
appropriate to the reduction to a seven-dimensional subspace. 

For example, the constraint function $G(x^\mu,p_\mu)=p_4$ gives the M2-brane phase space of Section~\ref{sec:M2flux} with the quasi-Poisson brackets \eqref{eq:M2brackets}. On the other hand, the choice $G(x^\mu,p_\mu)=x^4$ gives the M-wave phase space of this section with the quasi-Poisson brackets \eqref{eq:rhobrackets} (under the identification $\ell_s^3\,R=e\,\rho$). The two phase spaces are just different seven-dimensional slices of the same eight-dimensional phase space $\hat\Mcal'$ and are related by Born reciprocity $(x_\mu,p^\mu)\mapsto (p_\mu, -x^\mu)$, which is realised here as a $Spin(7)$-transformation. Other linear seven-dimensional slices of the eight-dimensional M-theory phase space $\hat\Mcal'$ are likewise related by rotations valued in $Spin(7)\subset SO(8)$.

\acknowledgments{
We would like to thank the organisors of the Workshop on ``Dualities
and Generalized Geometries'' for the invitation
to deliver these lectures, and all the participants for their
questions and comments which have helped shape the
structure and content of this article. We are particularly grateful to
David Berman, Bernd Henschenmacher, Jeffrey Heninger, Chris Hull, Vladislav Kupriyanov,
Dieter L\"ust, Nick Mavromatos and Philip Morrison for
discussions and correspondence on the contents of this paper.
This work was supported in part by the Action MP1405 QSPACE ~from the European Cooperation in Science and Technology
(COST) and by the Consolidated Grant ST/P000363/1
from the UK Science and Technology Facilities Council (STFC). 
}

\end{document}